\documentclass{article}

\usepackage[english]{babel}
\usepackage[letterpaper,top=2cm,bottom=2cm,left=3cm,right=3cm,marginparwidth=1.75cm]{geometry}
\usepackage{amsmath}
\usepackage{graphicx}
\usepackage[colorlinks=true, allcolors=blue]{hyperref}
\usepackage[capitalise,noabbrev]{cleveref}
\usepackage{mdframed}
\usepackage{subcaption}
\usepackage{multirow}
\usepackage{listings}
\usepackage{booktabs}
\usepackage{xcolor}
\usepackage{textcomp}
\usepackage{algorithm}

\date{}
\title{Revealing the influence of participant failures on model quality in cross-silo Federated Learning}
\author{Fabian Stricker -- fabian.stricker@h-ka.de \\ 
David Bermbach -- db@3s.tu-berlin.de \\
Christian Zirpins -- christian.zirpins@h-ka.de}

\begin{document}

\maketitle

\begin{abstract}
Federated Learning (FL) is a paradigm for training machine learning (ML) models in collaborative settings while preserving participants' privacy by keeping raw data local.
A key requirement for the use of FL in production is reliability, as insufficient reliability can compromise the validity, stability, and reproducibility of learning outcomes.
FL inherently operates as a distributed system and is therefore susceptible to crash failures, network partitioning, and other fault scenarios.
Despite this, the impact of such failures on FL outcomes has not yet been studied systematically.

In this paper, we address this gap by investigating the impact of missing participants in FL.
To this end, we conduct extensive experiments on image, tabular, and time-series data and analyze how the absence of participants affects model performance, taking into account influencing factors such as data skewness, different availability patterns, and model architectures.
Furthermore, we examine scenario-specific aspects, including the utility of the global model for missing participants. 
Our experiments provide detailed insights into the effects of various influencing factors. 
In particular, we show that data skewness has a strong impact, often leading to overly optimistic model evaluations and, in some cases, even altering the effects of other influencing factors.
\end{abstract}

\section{Introduction} 
Today, many companies employ machine learning (ML) models in their digital products, services, or processes.
The quality of these models, however, strongly depends on the availability of suitable training data, including data covering rare edge cases.
In practice, such data are often confined to isolated silos and cannot be shared due to corporate policies or privacy regulations~\cite{paper_pallas_fog4privacy}.

This problem is addressed through federated learning (FL), in which ML models are trained collaboratively without exchanging raw data~\cite{mcmahanCommunicationEfficientLearningDeep2017}.
To this end, multiple participants each train on local data and send the resulting ML model to a coordinator.
The coordinator then aggregates the partial models, e.g., based on weighted averaging of their parameters (FedAvg~\cite{mcmahanCommunicationEfficientLearningDeep2017}), and returns the resulting global model to all participants.
This is repeated over multiple rounds with evaluation and additional training until partial and global models converge~\cite{mcmahanCommunicationEfficientLearningDeep2017, strickerFLAPUSoftwareArchitecture2024a}.

As an FL deployment is inherently a distributed system, it is exposed to the full spectrum of failures typical of such systems.
Among others, FL deployments often have to deal with participants dropping out either temporarily or permanently.
In cross-device FL with many participants, e.g., based on mobile phones, this is the default state but can be addressed:
Due to the large number of participants, cross-device FL relies on participant selection methods and can easily fall back to alternative candidates~\cite{hardFederatedLearningMobile2019,riberoFederatedLearningIntermittent2023a,wangFriendsHelpSaving2024}.
In contrast, cross-silo FL includes just a few participants, each contributing a large share of the overall data~\cite[p.4-16]{kairouzAdvancesOpenProblems2021}.
Participants in cross-silo FL might drop out through various kinds of failures, e.g., Internet problems, software bugs, or energy outages.
In such cases, a failed participant cannot be replaced by another due to their small number and the likelihood of unique individual training data that is non-independent non-identically distributed (non-IID)~\cite{strickerAnalyzingImpactParticipant2025a}.
While there is work on fault tolerance in cross-device FL, e.g.,~\cite{riberoFederatedLearningIntermittent2023a, sousaEnhancingRobustnessFederated2025, imteajSurveyFederatedLearning2022}, this topic has received little attention in cross-silo FL with non-IID data~\cite{zhaoFederatedLearningNonIID2018}.

In this paper, we further address this gap by extending our previous work~\cite{strickerAnalyzingImpactParticipant2025a}, which explored the impact of participant failures in cross-silo FL on the training and evaluation of convolutional neural networks (CNNs) with image data, considering different influential factors. In addition, we analyzed how the performance deteriorates for the missing participant and the use of Shapley values (SVs) for impact detection. 

This paper considerably extends our former work by considering important additional data modalities such as tabular and time-series data that exhibit unique properties and structure and are often used in organizational contexts.
We also add a crash-recovery scenario to understand the impact of a participant failing and recovering near the end of the training process.
To further understand the limits of the influential factors, we simulated skew situations to consider highly unique heterogeneous data silos.
Lastly, we analyze the impact of different model architectures for image and time-series data to support architectural design choices.

Overall, this extended paper provides the following contributions:
\begin{itemize}
    \item We systematically analyze the problem of participant failures in cross-silo FL and derive failure-impact modifiers that are essential to understand the impact of failures. Furthermore, we provide a terminology of failure-impact modifiers (\cref{sec:influencing_factors}). 
    \item We conduct a comprehensive study across time-series, image, and tabular datasets while considering different non-IID data skews and additional failure-impact modifiers, while providing insights into the effect of different failure-impact modifiers on the training, evaluation, and the utility of the model for the missing participant. We make all data publicly available (\cref{sec:exp_plan}). 
    \item We discuss the findings and provide guidance for assessing the impact of participant failures (\cref{sec:findings}).
\end{itemize}

\section{Background \& Related Work}\label{sec:background}
In this section, we explain the concepts of FL including the structure of a FL system, characteristics of the participants, and data heterogeneity. Furthermore, we provide an overview of the related work.

\subsection{Concepts of Federated Learning}
\label{sec:fl_concepts}
Since FL is a new technology, there is not yet a common conceptual framework; instead, FL offers various patterns that fit diverse application contexts.

The first aspect to consider is the structure of the FL system, which can range from centralized to fully decentralized.
In centralized FL, there is only one coordinator instance that handles all participants and aggregates the weights.
This approach is often used because of the centralized orchestration of the process, but it is susceptible to single points of failure and may lead to a bottleneck. 
In decentralized FL, the coordinator role is handled by all participants individually, i.e., each participant also aggregates the models~\cite[p.11-13]{kairouzAdvancesOpenProblems2021}.
In between are different architectures such as multiple aggregation layers to handle large numbers of participants or multiple coordinators to improve robustness~\cite{bonawitzFederatedLearningScale2019}.
Next, the data structure is important as it leads to different training processes. 
In particular, data can be partitioned horizontally or vertically.
If data are horizontally partitioned (HFL), then all participants have the same features for a sample and can train on their datasets.
Contrary, in vertical FL (VFL), a single participant does not have all features for a sample and therefore requires communication with the other participants for training~\cite{Ludwig2022}.
The next aspect concerns aggregation methods, as they define how the model weights of each participant are combined and hence influence the learned patterns. 
The most common method is FedAvg~\cite{mcmahanCommunicationEfficientLearningDeep2017}, but many new and modified algorithms emerged to improve different aspects such as robustness, privacy, or quality~\cite{gargClientAvailabilityFederated2025, liuPrivacyPreservingAggregationFederated2022}.
The last aspect addresses the characteristics of the FL participants, which are commonly distinguished into two settings.
The first setting is called cross-\emph{device} FL and consists of many participants that are mostly mobile devices, which are constrained by limited internet connectivity, battery capacity, and usually have little data samples~\cite{bonawitzFederatedLearningScale2019}.
In contrast, the second setting is called cross-\emph{silo} FL and consists of a few participants, where each participant has substantial resources and large datasets.

In real-world FL systems, data skewness is a common issue that negatively impacts the quality of the model and can manifest in different, potentially overlapping types~\cite{zhaoFederatedLearningNonIID2018}.  
The most common is quantity skew, where participants have different numbers of data samples. 
In label skew, the label distribution differs across participants. 
Finally, with attribute skew (full feature overlap), participants share the same feature space, but the feature distributions differ~\cite{zhuFederatedLearningNonIID2021}.
Depending on the data modality, other skews can occur.
For example, Nachtigall et al.~\cite{nachtigallRevealingEffectsData2024} reported new skews for time-series datasets such as seasonality skew, and McMahan et al.~\cite{mcmahanCommunicationEfficientLearningDeep2017} found that non-IID data may cause model divergence leading to unstable convergence.

In the following, we focus on cross-silo FL, where the number of participants is small, and each participant holds a significant portion of the overall data. In such settings, participant failures can have a substantial impact on the training process and the resulting model quality. Therefore, we next review related work that addresses participant failures in FL.

\subsection{Related Work}
\label{sec:related_work}

In existing research, the impact of participant failures in cross-\emph{silo} FL is often overlooked, and comprehensive analyses are missing. 
On the contrary, research on missing participants in cross-\emph{device} FL is more mature, as participants not being able to participate are common due to their sheer number. 
For the discussion of related work, we start with existing work on missing participants in cross-device FL.
Here, one approach to mitigate the issue of a missing or unavailable participant is to use modified participant selection methods.
Sousa et al.~\cite{sousaEnhancingRobustnessFederated2025} propose an extension for selection strategies that, in case of a failure, selects a different participant from a predefined reserve pool of substitute participants to ensure convergence of the global model.
However, using a reserve pool only is applicable if there are sufficient participants.
Another aspect is the bias caused by participants with higher availability.
Here, Ribero et al.~\cite{riberoFederatedLearningIntermittent2023a} proposed an unbiased selection strategy that includes the availability of participants to prevent bias.
Besides participant selection strategies, there are also substitution and augmentation strategies to tackle convergence issues~\cite{xuStabilizingImprovingFederated2024}.
Wang and Xu~\cite{wangFriendsHelpSaving2024} use the data distribution to identify similar participants. In case of a failure, the model update of the similar participant is used as a substitute to reduce the impact. Their approach is based on similarity and clustering, which is difficult to achieve with few participants and skewed data.
Instead of using a similar participant as a substitute, Sun et al.~\cite{sunMimiCCombatingClient2024} modify model updates with a correction value to mitigate convergence issues in mobile edge networks.
Besides mitigating the issues, there are also works that focus on system threats like dropout attacks causing system disruptions.
Qian et al.~\cite{qianDROPFLClientDropout2024} propose an attack under constraints such as low bandwidth and unstable network connection. They found that attacking a high value participant slows down convergence and has more influence in a non-IID setting. Following this work, assessing the contribution of participants could help in detecting the impact if specific participants crash.

Although focusing on cross-device instead of cross-silo FL, some existing studies share closer similarities with our work.
Huang et al.~\cite{huangKeepItSimple2023} analyze the impact of unreliable devices in a rural environment and consider non-IID data in combination with different participation rates. Their results indicate that FedAvg is robust against unreliable participants; however, their data contain only minor skews and few classes.
Ruan et al.~\cite{ruanFlexibleDeviceParticipation2021} analyze flexible device participation, where participants can send incomplete updates, join or leave the process.
They focus on reducing the bias of incomplete updates as well as how new participants may accelerate the training.
While they consider a cross-device scenario, they provide an analysis on how dynamic participation influences the convergence.
So far, aspects such as convergence, bias, and the availability of participants are considered, but mostly for cross-device scenarios and the training process.
To the best of our knowledge, a detailed analysis for cross-silo FL that also includes a range of additional factors such as the model architecture, complexity of the dataset, impact on the evaluation, and different number of participants is still missing.
In our former work~\cite{strickerAnalyzingImpactParticipant2025a} we have provided first insights into the effects of participant failures in cross-silo FL for different factors while focusing on image datasets with CNN models. 
In particular, we have focused on analyzing the impact on the evaluation, the performance difference of the quality for the missing participants against the other participants, and the use of SVs to predict the impact.

In this paper, we analyze new aspects that were not addressed yet.
First, we study different data modalities, such as tabular data, which are commonly used in organizational settings as well as different model architectures.
We also analyze the impact of unique heterogeneous silos, which represent an extremely skewed distribution, to show the limitations of the influential factors.
In addition, we discuss crash-recovery behavior to analyze the influence of a participant rejoining the process.

\section{Participant Failures and Impact Modifiers}\label{sec:influencing_factors}
In this section, we give an overview of participants failures in cross-silo FL and analyze which potential factors can alter the impact on the performance of the model.

\subsection{Participant failures in FL systems}
In FL, failures can occur on the side of the coordinators (i.e., crashing coordinator, wrong FL configuration) or on the side of the participants. Here, we focus on the participant-related failures.
To categorize the failures, we use their natural characteristics.
These include the duration of the failure, the timing of the failure, the failure type, and the cause of the failure. 
Usually the duration is measured in time; however, in some FL phases (i.e., training, evaluation), the rounds can be used as a measure.
For the timing, we can use the FL phases, such as training and evaluation, to indicate when the failure occurs. 
Last, for failure types, we consider the common failures such as crash, communication, or timing failures.

In FL, the most commonly reported failures are participant crashes, network failures, participants that require significantly more time for computationally expensive tasks (i.e., model training), and Byzantine failures sorted by their detectability~\cite{wangFLuIDMitigatingStragglers2023, allouahByzantineRobustFederatedLearning2024}.
Participant crashes can occur during any phase of the FL process and can be caused by software bugs or human-made mistakes. 
Here, the failure type is a crash failure. 
The duration depends on the cause and the recovery, as a software bug could lead to another crash after recovery.
A participant crash can influence the assessment and the quality of the model, as the participant's data are missing during the training or evaluation phase.
As a consequence, unique data can be missing, which limits the patterns learned by the model.

A similar behavior can occur with network failure, such as an external link failure that persists for a longer period.
In this case, the participant cannot reach the coordinator and, therefore, is unable to send results to the server.
Here the duration depends on the infrastructure, short interrupts can be resolved with reliable communication, while longer interruptions can lead to missing updates for multiple rounds.
An important difference is that the participant can still perform operations.

The next failure is slow participants that send the results of expensive operations after a deadline is met, which is a timing failure.
In cross-silo FL, some participants can have more data than others, which can lead to larger differences between the individual training times.
Here, the issue only occurs during computationally expensive operations such as the training.
Furthermore, the impact on the FL system is limited as it depends on the aggregation method.
Some systems would wait until the participant is finished, while others define a deadline, which would lead to missing updates.

The last failure is a Byzantine failure, where the participant could send wrong information, such as wrong training or evaluation results. 
Here, communication issues, software bugs, or wrong configuration of the system could lead to a different training configuration than the one provided by the coordinator.
Depending on the cause, the failure can persist for a longer duration or happen once for a shorter period.
For example, the participant could calculate the accuracy, but precision was supposed to be calculated. 
In FL, this issue can lead to confusing and wrong results or even to a malfunction of the process.

For the remainder of this work, we define a missing participant as a participant who, during a failure period, is unable to execute any task given by the coordinator.
Here, we consider participant crashes and link failures with a long duration that occur during the training and evaluation phase.
For brevity, we use failure to denote participant failure.
Next, we will analyze impact modifiers that increase the impact on the FL system.

\subsection{Failure-impact modifiers}
In recent work on missing participants, the aggregation method, availability, and non-IID data have been mentioned as factors that influence the impact of a failure~\cite{qianDROPFLClientDropout2024,sunMimiCCombatingClient2024,huangKeepItSimple2023}.
Here, each of them modifies either the behavior or the results of the participant.
The first modifier, the aggregation method, defines how the results are combined and the participant use that result for the next training. The aggregation method is part of the FL process design.
Next, the availability is connected to the characteristics of failures and describes when data are available.
The last modifier influences the data distribution across the participants, which can lead to silos that contain unique information.
Consequently, the modifier influences the result of the participant and the overall data distribution if a participant crashes.

Following these aspects, we analyze the overall FL process that includes the role of data, the design of the process, and the role of the participants, as well as the characteristics of the failure to identify additional aspects that influence the impact of a missing participant.
We call these aspects failure-impact modifiers (FIM) and categorize them into data-specific, design-specific, and coordination-specific. An overview of all FIMs that are used in the experiments is provided in \cref{tab:configuration_range}. In the following, we discuss the modifiers in each category: 

\textbf{Data-specific Modifiers} 
The data-specific modifiers include all aspects that manipulate the data of the participant.
Here, the identified modifiers in the FL process include the data skewness, dataset complexity, data modalities, and data-record partitioning.
The first modifier is the data modalities (DMod), which can lead to dependencies between features or samples such as in tabular or time-series data.
Furthermore, it indirectly influences the selection of the model architecture, training hyperparameters (i.e, loss, evaluation metrics) and data skewness.
Second, the data skewness (DSkew) is a failure-modifier, as a higher skew (both quantity and label skew) can lead to participants with unique data.
Hence, specific data are only contributed by single silos, which, in the case of a dropout, would lead to their missing (i.e., unique classes or edge cases)~\cite[p.18-20]{kairouzAdvancesOpenProblems2021}.

The dataset complexity (DComp) indicates how difficult it is for a model to extract meaningful patterns from the data. 
This can change due to factors such as the number of unique classes, feature dimensionality, and quality of the data (e.g., meaningful features, missing values, reflection of real-world population).
For example, high complexity requires more samples to extract and learn complex patterns.
If a participant crashes and recovers later, then a lower complexity can yield a faster recovery of the model performance.
The last modifier is the data-record partition (DPartition) which refers to VFL and HFL.
The impact of this factor follows the different available features across the participants.
In VFL, the impact is higher, as a missing participant would lead to being able to train the model or make predictions~\cite{Ludwig2022}.
In contrast, in HFL a missing participant in the FL process can still continue as each participant has the full features for a sample.

\textbf{Design-specific Modifiers}
The design-specific modifiers are related to the design of the FL process and consider aspects that modify the behavior of the participants.
First, the model architecture (MArch) can influence the impact of a failure through different complexities of the model as well as different kinds of layers as they influence how the model learns patterns.
For example, if the model consists of many generalization layers, then it tries to generalize rather than focusing on the specific data of the participants.
Next, the training hyperparameters (HParams) (i.e., learning rate, optimizer) can influence the impact of a missing participant because they define how the model learns. 
For example, a higher learning rate leads to large steps and can lead to faster forgetting of information after a failure.
The next modifier is the aggregation method (Amethod).
Similar to the hyperparameters, the aggregation method influences the weights, but through the aggregation of the updates on the coordinator side. If a participant fails, then, depending on the aggregation method, the participants could have different weights assigned, which leads to a different influence on the global model.
The last modifier is the number of participants (NParti).
The number of participants can influence the impact of a missing participant, as usually more data means more different distributions, which can lead to better generalization.
Hence, more participants could lead to less impact, while fewer participants can lead to a higher impact because less data are available and the impact of a participant is more significant on the data.

\textbf{Coordination-specific Modifiers} 
This category covers coordination-related aspects of the FL system, including tasks that are delegated to the participants and failure characteristics that affect coordination.
The first modifier is the participant task (PTask). 
In general, the task influences the impact, as in FL the different tasks lead to different results.
These tasks include the data preprocessing (PTask.DPrep), training (PTask.Train) or evaluation (PTask.Eval).
If a participant fails during training or evaluation, then the result will be missing, which can lead to a different resulting global model or to different evaluation results.
The last modifiers based on the failure characteristics are the duration (FChar.Dur) and the timing, which lead to different availability phases for the missing participant (FChar.Ti).
If a participant crashes, the duration is relevant because, depending on it, more updates can be missing, which leads to a larger shift in the global model.
Next, failures can occur in random patterns, leading to different availability phases during which a participant's data are included in the process.
If the participant contributes early but crashes, his information can be forgotten over the training process.
In contrast, a recovery and participation in the later rounds can refresh the information.
Based on the derived failure-impact modifiers, we will next define the experiment plan and the research goals to analyze the impact.

\section{Experiment Plan}\label{sec:exp_plan}
In this section, we formulate the research goals and research questions of the experimentation study.
Furthermore, we describe the setup of the experiments and how they are conducted.

\subsection{Experimentation Goals and Research Questions}
\label{sec:experiment_goals}

The purpose of this study is to explore the effects of the FIMs in the failure scenario discussed in \cref{sec:influencing_factors} on the quality of the model in cross-silo FL scenarios with few participants. 
Covering all modifiers in this study is not feasible due to the vast number of experiments; hence, we will explore a subset of modifiers.
Furthermore, analyzing the influence of the modifier alone is not possible, as they need to be connected to a task such as evaluation or training where the impact can be measured.
Besides the modifiers, we also analyze scenario-relevant influences, as these help in understanding the consequences for cross-silo FL.
Here, one important aspect is fairness across the participants; hence, the influence on the usability of the model for the missing participants needs to considered.
Based on the modifiers and the scenario, we formulate the following research questions:

\noindent\textbf{RQ1} (Related to FIMs: DMod, DSkew, DComp, MArch, PTask.Eval) \emph{Can a missing participant influence the model evaluation and, therefore, lead to wrong assumptions about the model quality?}
The evaluation is essential to measure the model quality and make deployment decisions.
Here, we aim to understand if a missing participant during the evaluation phase can influence the result (i.e., the metrics) and whether a resulting difference is significant enough to lead to a biased evaluation.
In this case, we consider the modifiers DMod, DSkew, DComp and MArch because they influence the properties of the data as well as the model, which are essential for making predictions.

\noindent\textbf{RQ2} (Related to FIMs: DMod, DSkew, DComp, MArch, FChar.Ti, PTask.Train)  \emph{How does a missing participant affect the training of the model and, therefore, the resulting quality?}
The goal of FL is to train a high-quality model; however, a missing participant leads to missing data, which is essential for training.
Here, the goal is to understand how a missing participant influences the quality.
In FL, the model is dependent on the previous training rounds. 
Therefore, the timing of a failure can influence the behavior of the model.
Furthermore, the data-specific modifiers and the modifier MArch are important as they influence the learning of patterns.

\noindent\textbf{RQ3} (Related to FIMs: DMod, DSkew, DIM.DC, NParti) \emph{How does a participant failure affect the usability of the trained model for the missing participant?}
In our setting, each organization should benefit from the process for them to keep participating. 
If the participant crashes and the model cannot be applied for a participant, then this influences the fairness of the process.
Therefore, the goal is to analyze the change in model performance for the learning task of the missing participant while considering the modifiers NParti, DMod, DSkew and DComp as they influence the data distribution and the overall available data.

\noindent\textbf{RQ4} (Contribution measures) \emph{Do participant contribution measures reflect the impact of a missing participant on the training task?}
In FL, participant contribution measures are often used as indicators for the importance of a participant.
If the impact of a missing participant can be predicted based on the importance, then this can help in designing intelligent fault-tolerant mechanisms.
The goal is to understand if the contribution measure reflects the impact of a failure on the training.

\subsection{Experiment Design}
In the experimentation goals (\cref{sec:experiment_goals}), we defined a set of research questions concerning different failure-impact modifiers.
To analyze the influence of each modifier, we define a base configuration (\cref{tab:base_configs}) that describes the default experiment setup for all datasets.
Further, we define the possible values a modifier can take in \cref{tab:configuration_range}.
We aim to isolate individual modifiers to analyze their effects without other influences.
Therefore, we will only change one modifier at a time.
However, there are still dependencies between these factors that cannot be resolved completely, e.g., the model architecture depends on the structure of the dataset.
For each research question, we will conduct a set of experiments based on the base configuration and the possible values of the modifiers to analyze the influence of each related modifier.
For \emph{RQ1}, we modify DMod, DComp, DSkew, and MArch, which results in a set of experiment configurations that analyze the influence of the participant included and excluded from the evaluation.
Similarly, for \emph{RQ2}, our objective is to understand the impact of the FIMs on the quality. 
However, some FIMs such as FChar.Ti are essential and cannot be removed, as their combination leads to the influence on the model quality.
Therefore, we analyze the impact of DMod, DComp, DSkew, and MArch while considering the FChar.Ti.
Next for \emph{RQ3}, we iterate over values of the modifiers DMod, DComp, DSkew, and NParti to analyze their impact on the usability of the global model for the missing participant.
Last, for \emph{RQ4}, we only consider the image datasets across the base skew and high skew for DSkew. Since SV computation requires a server-side dataset, the distributions for both datasets are reduced by a fixed quantity per class. 
In the following, we describe the experiment components in detail.
\begin{table}[t!]
\centering
\caption{Base configuration for all datasets}
\label{tab:base_configs}
\begin{tabular}{@{}ll@{}}
\toprule
\textbf{Parameter} & \textbf{Value} \\
\midrule
Local Rounds, Global Rounds & 5 Rounds, 30 Rounds \\
Aggregation Strategy, Optimizer (Learning rate)        & FedAvg, Adam (0.0001)       \\
Availability Timing, Available Rounds & Start, 2  \\
Metrics (Classification, Forecasting) & Macro F1-score, R2-score \\
\bottomrule
\end{tabular}
\end{table}

\begin{table}[t!]
\centering
\caption{Overview of all modifiers used in the experiments with description and possible values. Default values are shown in \textbf{bold}}
\label{tab:configuration_range}
\begin{tabular}{@{}lll@{}}
\toprule
\textbf{Modifiers} & \textbf{Description} & \textbf{Range of values}  \\
\midrule
DSkew   & Data Skew & \textbf{Base Skew}, High Skew, Manual Skew \\
NParti   & Number of Participants & \textbf{4}, 6, 8 \\
FChar.Ti & Participant Availability & \textbf{Start}, Middle, End, StartEnd  \\ 
DMod   & Data Modality & \textbf{Image}, Tabular, Time-series \\
MArch (Image) & Model Architecture & \textbf{CNN}, ResNet50 \\ 
MArch (GermanSolarFarm) & & \textbf{LSTM}, 1D-CNN \\
DComp (Image) & Dataset Complexity & \textbf{CIFAR10}, CIFAR100 \\    
DComp (Tabular) & & \textbf{Adult}, Covertype \\ 
DComp (Time-series) & & \textbf{PVOD}, GermanSolarFarm \\
PTask & Participant Task & Training, Evaluation  \\
\bottomrule
\end{tabular}
\end{table}

\subsubsection{Datasets and Partitioning}
For the experiments, we use \textit{CIFAR10}~\cite{krizhevskyLearningMultipleLayers2009a} and \textit{CIFAR100}~\cite{krizhevskyLearningMultipleLayers2009a} as image datasets for classification.
The datasets have a similar structure but vary in the number of classes, which leads to a complexity difference that we can leverage. 
Each dataset consists of 60,000 data samples that are split equally across the number of classes.
As tabular datasets, we use \textit{Adult}~\cite{beckerAdult1996} and \textit{Covertype}~\cite{blackardCovertype1998} because they contain class imbalance and differ in complexity.
\textit{Adult} consists of 48,842 data samples with 14 features and two classes.
In comparison, \textit{Covertype} consists of 581,012 data samples with 54 features and seven classes. 
Last, \textit{PVOD}~\cite{yaoPVODV10Photovoltaic2021} and \textit{GermanSolarFarm}~\cite{genslerGermanSolarFarmDataSet2016} are time-series datasets for regression. 
Both consist of multiple datasets collected by photovoltaic (PV) stations with different locations. 
The \textit{PVOD} dataset has 14 features with 15-min resolution.
For the experiments, stations act as one FL participant each and are selected by the station number.
In comparison, \textit{GermanSolarFarm} consists of 50 features with a resolution of three hours.
For the experiments, we assign the stations non-uniformly to participants to simulate utility companies of different sizes (few vs. many PV sites).
We partition the datasets into non-IID datasets by using the Dirichlet distribution~\cite{linEnsembleDistillationRobust2020}.
Here, we sample the quantity for eight participants and afterward calculate the label distribution bounded to the quantity skew to ensure that no data record is duplicated.
For the experiments, we create the following skews:
\begin{itemize}
    \item \textit{Base skew} (BS) consists of a quantity skew alpha of 1.5 and an alpha value of 0.8 for the label skew.
    \item \textit{High skew} (HS) consists of a quantity skew alpha of 1.2 and an alpha value of 0.6 for the label skew, which aims to provide more insights into the impact of the distribution.
    \item \textit{Manual Skew} (MS) consists of data partitions in which participants share some classes while retaining unique ones to simulate organizations with partially overlapping distributions and distinct edge cases.
\end{itemize}

\subsubsection{Model Architecture}
For both image datasets, we use CNNs because they achieve good performance while the training duration is short.
In addition, we also use \textit{ResNet50}~\cite{heDeepResidualLearning2016}, a widely used deep residual CNN architecture, to analyze whether the model architecture influences the impact of a failure.
In the case of the tabular datasets, we use a \textit{TabTransformer}~\cite{huangTabTransformerTabularData2020} because it achieves high performance.
For \textit{PVOD}, we use a 1D-CNN which is efficient while also achieving good performance. For the \textit{GermanSolarFarm} dataset, we use a long short-term memory (LSTM) model and a 1D-CNN with increased dilation for analyzing the impact of different architectures.
For the training of the models, we use a train-test split with a ratio of 70/30.
The implementation of the models is publicly available\footnote{Code available at: \url{https://github.com/HKA-IDSS/Supplement-Revealing-the-influence}}.

\subsubsection{Contribution Measurement}
For quantifying the contribution of a participant, we use Shapley-based methods because they are accurate.
However, they also cause high computational overhead~\cite{wangPrincipledApproachData2020}.
In the experiments, we use the \textit{RoundSVEstimation} method based on permutation sampling proposed by Wang et al.~\cite{wangPrincipledApproachData2020}. While the method requires a coordinator-side dataset, it mitigates the high computational overhead.
Furthermore, the method estimates the contribution at each round; hence, it is suited to make conclusions in a round before the failure happens.

\subsubsection{Simulating Participant Failures}
To simulate a missing participant, we decide, based on the availability phase, that a single participant is unable to perform training or evaluation tasks. We assume that participant 0 fails and will participate only during the availability phases.
As availability timing options, we consider the \texttt{start} of the process (i.e, the first two rounds), the \texttt{middle} of the process, or the \texttt{end} (i.e., the last two rounds).
For the timing \texttt{startEnd} (failure with crash-recovery), the participant will participate in the first two rounds and then rejoin in the last two rounds.
This way we can analyze the timing effects of a participant leaving and optionally rejoining.

\subsubsection{Performance Evaluation}
To assess the model performance, we leverage the participants test datasets and aggregate their evaluation results.
For classification tasks, we send the confusion matrix to the coordinator and calculate the metrics based on combined confusion matrix.
In our scenario, the unique classes of participants are important and should be learned by the model to improve the overall quality.
Therefore, we use the macro F1-score as it represents the behavior and can handle class imbalance.
For the time-series datasets, we use R2-score which represents the ability of the model to explain the variability in the data.

\section{Results}\label{sec:findings}
This section presents the experimental results and provides a structured overview of the findings for each research question.
%Furthermore, we discuss aspects that may influence the validity of the findings.

% \subsection{Findings}
% In this subsection, we provide structured answers for each research question and highlight the findings obtained from the experiments.

\subsection{RQ1: Can a missing participant influence the model evaluation and, therefore, lead to wrong assumptions about the model quality? }
\begin{mdframed}
\textbf{Experiments:} We compare the difference between evaluation including and excluding the missing participant to understand the impact of the missing information while considering DMod, DSkew, DComp, and MArch.

\noindent\textbf{Finding:} \emph{A participant missing during evaluation leads to an overly optimistic evaluation across different data modalities.}
Here, modifier DSkew influences the impact strongly, while a higher skew leads to a higher impact on the evaluation. 
Furthermore, the label skew shows more influence than the quantity skew.
For modifier DComp, a low complexity leads to a reduced impact.
The highest impact was for datasets that have many samples and many classes that enable a more skewed distribution.
Regarding MArch, we could observe that some architectures are more robust than others depending on the generalization layers as well as how information is retained.
\end{mdframed}

\begin{figure}[t]
    \centering
    \begin{subfigure}{0.48\textwidth}
        \centering
        \includegraphics[width=\linewidth]{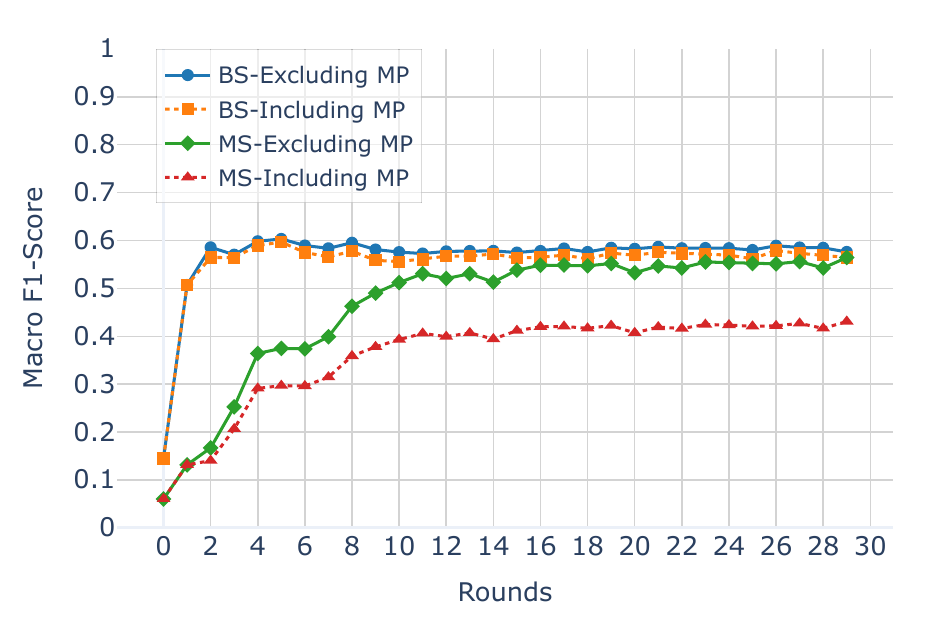}
        \caption{CIFAR10}
        \label{fig:cifar10_eval}
    \end{subfigure}
    \begin{subfigure}{0.48\textwidth}
        \centering
        \includegraphics[width=\linewidth]{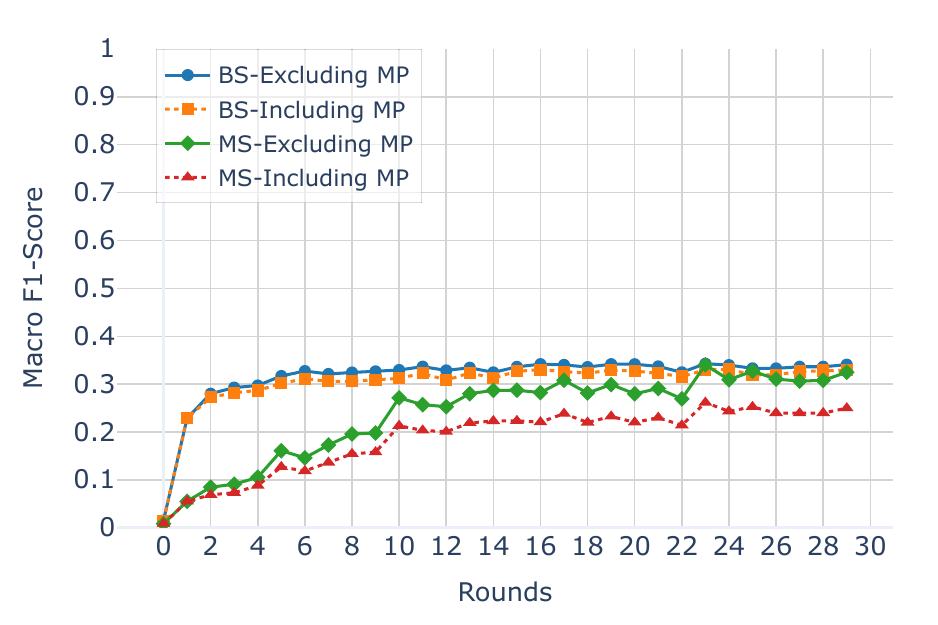}
        \caption{CIFAR100}
        \label{fig:cifar100_eval}
    \end{subfigure}
    \caption{Image Datasets: Comparison of the evaluation with and without the failed participant across base skew (BS) and manual skew (MS)}
    \label{fig:cifar10_100_eval_skews}
\end{figure}

\begin{figure}[t!]
    \centering
    \begin{subfigure}{0.48\textwidth}
        \centering
        \includegraphics[width=\linewidth]{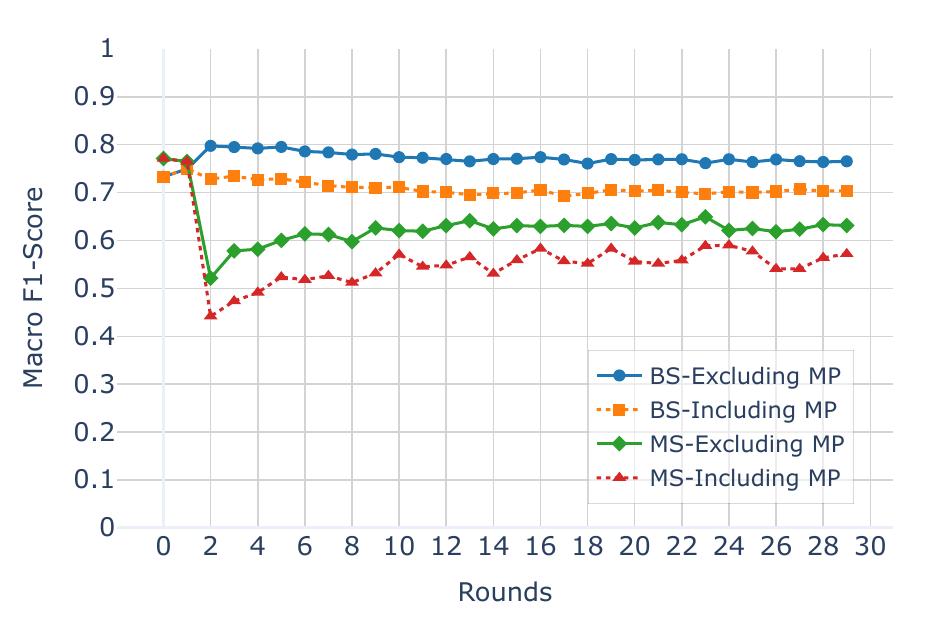}
        \caption{Adult}
        \label{fig:adult_eval}
    \end{subfigure}
    \begin{subfigure}{0.48\textwidth}
        \centering
        \includegraphics[width=\linewidth]{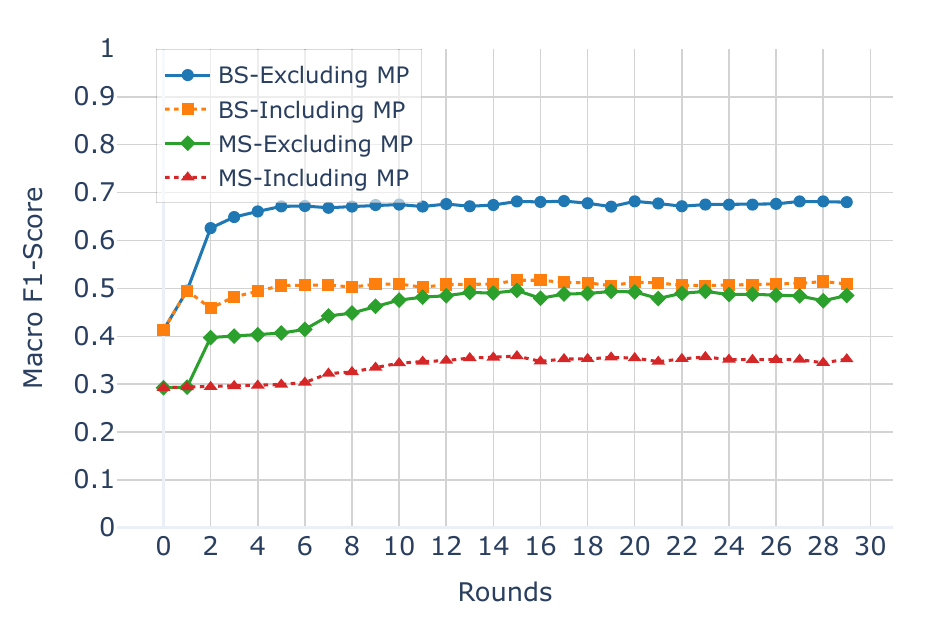}
        \caption{Covertype}
        \label{fig:covertype_eval}
    \end{subfigure}
    \caption{Tabular Datasets: Comparison of the evaluation with and without the failed participant for the BS and MS}
    \label{fig:adult_impact_evaluation}
\end{figure}

\noindent We start with analyzing \textit{CIFAR10} in Fig.~\ref{fig:cifar10_eval} and can identify that an evaluation excluding the missing participant leads to a higher F1-score than using all participants for both skews.
In the case of the manual skew, where the samples are equally distributed, but each silo contains a unique class, we can identify a higher impact. 
Here, the label skew has a higher relevance than the quantity skew. 
The \textit{CIFAR100} dataset with different skews in Fig.~\ref{fig:cifar100_eval} shows that the effect is less pronounced compared to \textit{CIFAR10}.
We assume that the key factor is the DComp as the high number of classes and the rather few samples per class in combination with the skew make it difficult to learn the patterns for all classes.
Furthermore, the \textit{CIFAR100} model includes multiple generalization layers. Therefore, the impact of the quantity per class can be lower as the model tries to generalize over all classes.
Hence, a large quantity of a specific class is less important.
Next, the manual skew in Fig.~\ref{fig:cifar100_eval} shows a strong change, which supports the assumption that DSkew has a high influence on the impact.
To understand how modifier FMI.DM affects the influence of a missing participant, we analyze the tabular datasets in Fig.~\ref{fig:adult_impact_evaluation}.
For both datasets, we can make the same observations as the image datasets for the modifier FMI.DS.
Contrary to the image datasets, the more complex dataset has a higher difference than the less complex dataset. 
Here, \textit{Adult} has only two classes; hence, having a high label skew is difficult, which is why the impact can be lower and other participants may supplement the information. Instead, \textit{Covertype} has more classes and can have a stronger skew.
Following, the FMI.DC shows high impact that is also connected to modifier FMI.DS.

\begin{figure}[t]
    \centering
    \begin{subfigure}{0.48\textwidth}
        \centering
        \includegraphics[width=\linewidth]{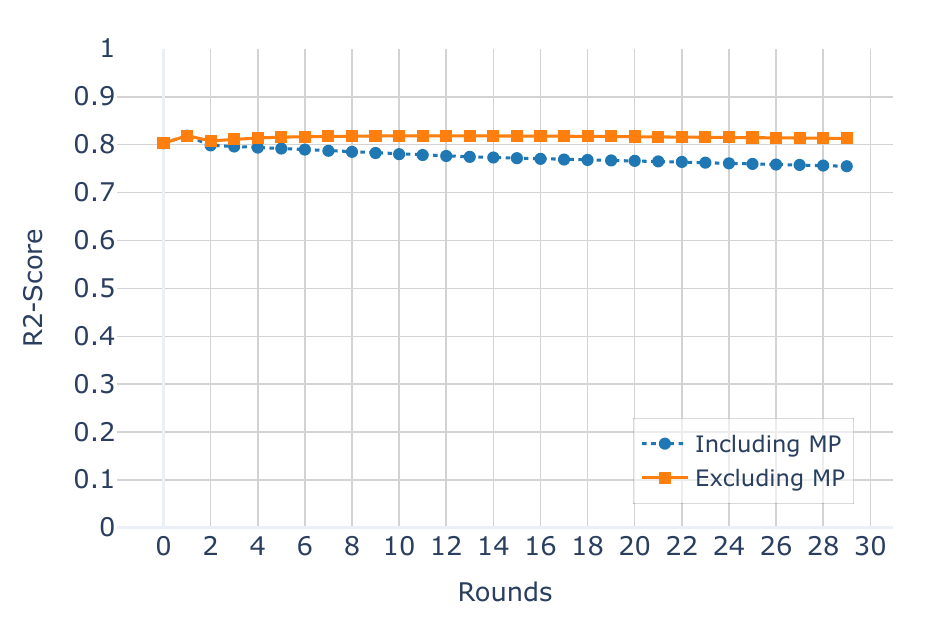}
        \caption{LSTM}
        \label{fig:germansolarfarm_eval}
    \end{subfigure}
    \begin{subfigure}{0.48\textwidth}
        \centering
        \includegraphics[width=\linewidth]{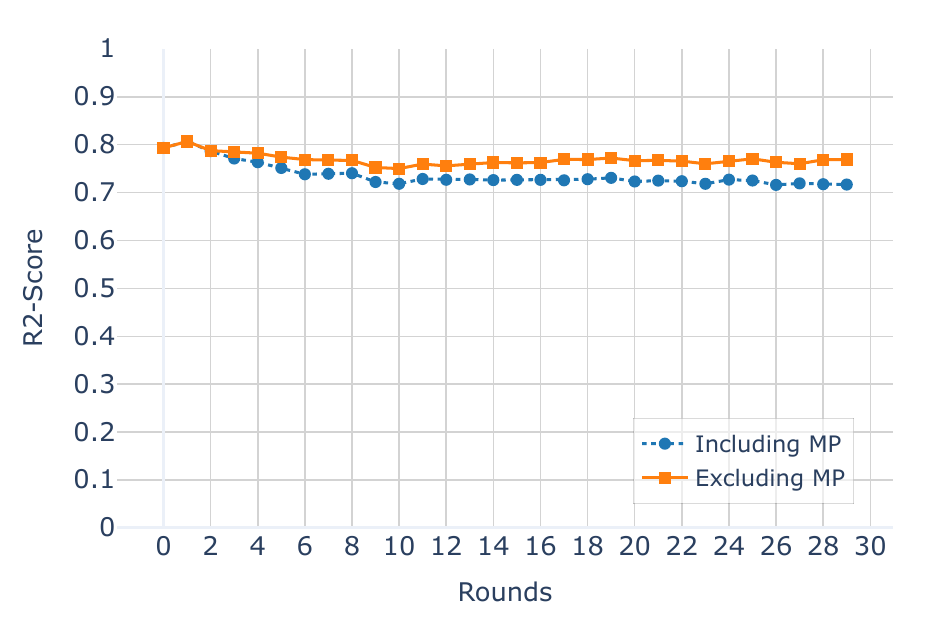}
        \caption{1D-CNN}
        \label{fig:germansolarfarm_cnn_eval}
    \end{subfigure}
    \caption{GermanSolarFarm: Comparison of the evaluation with and without the missing participant across different models}
    \label{fig:germansolarfarm_impact_evaluation}
\end{figure}

The last data modality are the time-series datasets with regression tasks. 
The \textit{GermanSolarFarm} shows the same behavior with a stronger effect in Fig.~\ref{fig:germansolarfarm_eval} that seems to increase over time.
We will omit the visualization for the \textit{PVOD} dataset, as it shows similar results with a lower impact.
To exclude the possibility that the states within the LSTM model could be the issue, we analyze the different model architectures in Fig.~\ref{fig:germansolarfarm_cnn_eval}.
We can observe the same behavior in both cases, while the LSTM appears to be more robust as the impact does not cause fluctuations in the quality compared to the 1D-CNN.
Following the experiments, the shift in test data, which excludes specific classes completely or reduces them to a bare minimum, can lead to a deviation between the testing goals.
In our scenario, this behavior is not wanted, as all companies should be able to use the model for their services.
The answer to this research question is important, as there are many methods that build upon utility measurements (e.g., early stopping) and hence, if a failure occurs, this can lead to unexpected behavior, which affects the reliability of a system.

\subsection{RQ2: How does a missing participant affect the training of the model and, therefore, the resulting quality?}
\begin{mdframed}
\textbf{Experiment:} We compare the performance of the different availability timings against the baseline without failures, while considering DMod, DSkew, DComp, MArch, and FChar.Ti.

\noindent\textbf{Finding:} \emph{A missing participant can lead to a decrease in model quality across different data modalities for scenarios with high skew.}
Here, it is important to understand that the modifiers also influence each other, which leads to the quality loss.
Across the experiments, we observed that FChar.Ti can affect the quality if the scenario does not consist of silos with extremely skewed classes.
In case of tabular datasets, the combination of DComp and DSkew influences the modifier FChar.Ti and can confuse model, which can degrade the performance.
In addition, DSkew has a high impact, especially for the manual skew with unique silos, where the influence of FChar.Ti is reduced.
Overall, a higher DSkew can lead to a higher performance loss across different DMods.
The MArch can influence the performance as the LSTMs show more robustness about the loss of quality over time.
The ResNet architecture showed similar behavior to the CNN architectures, while the overall training process fluctuated more. 
\end{mdframed}

\begin{figure}[t]
    \centering
    \begin{subfigure}{0.49\textwidth}
        \centering
        \includegraphics[width=\linewidth]{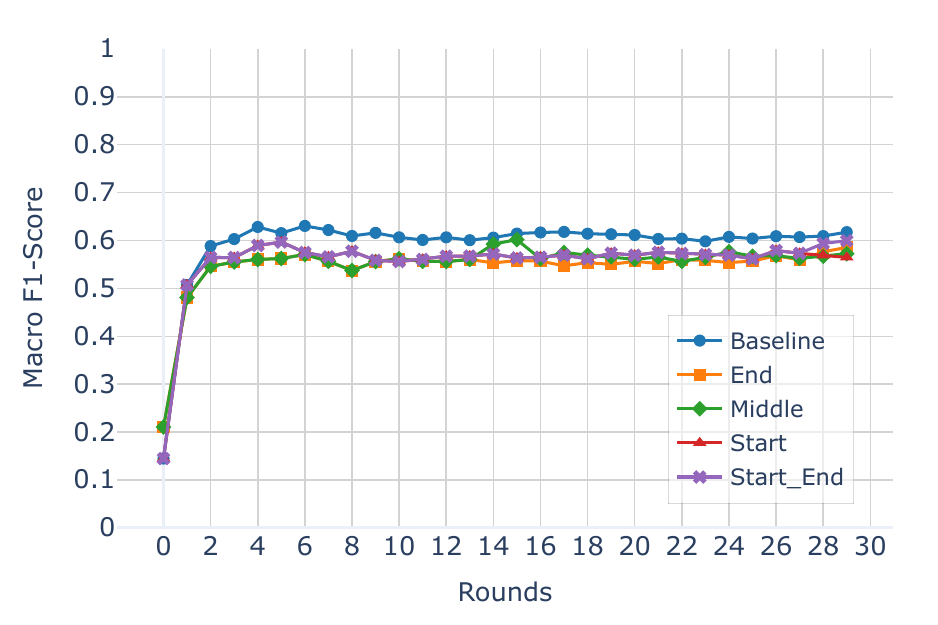}
        \caption{Base Skew}
        \label{fig:availability_impact_cifar10_bs}
    \end{subfigure}
    \begin{subfigure}{0.49\textwidth}
        \centering
        \includegraphics[width=\linewidth]{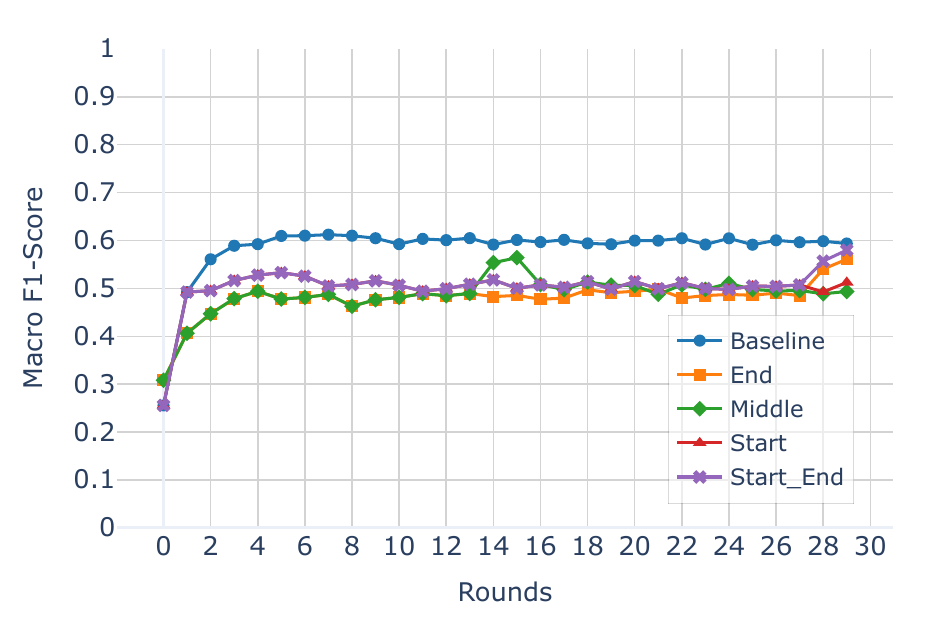}
        \caption{High Skew}
        \label{fig:availability_impact_cifar10_hs}
    \end{subfigure}
    \caption{CIFAR10 CNN: Different availability phases for the missing participant across different skews. The baseline represents the absence of the failure}
    \label{fig:cifar10_timing}
\end{figure}

\begin{figure}[t]
    \centering
    \begin{subfigure}{0.48\textwidth}
        \centering
         \includegraphics[width=\linewidth]{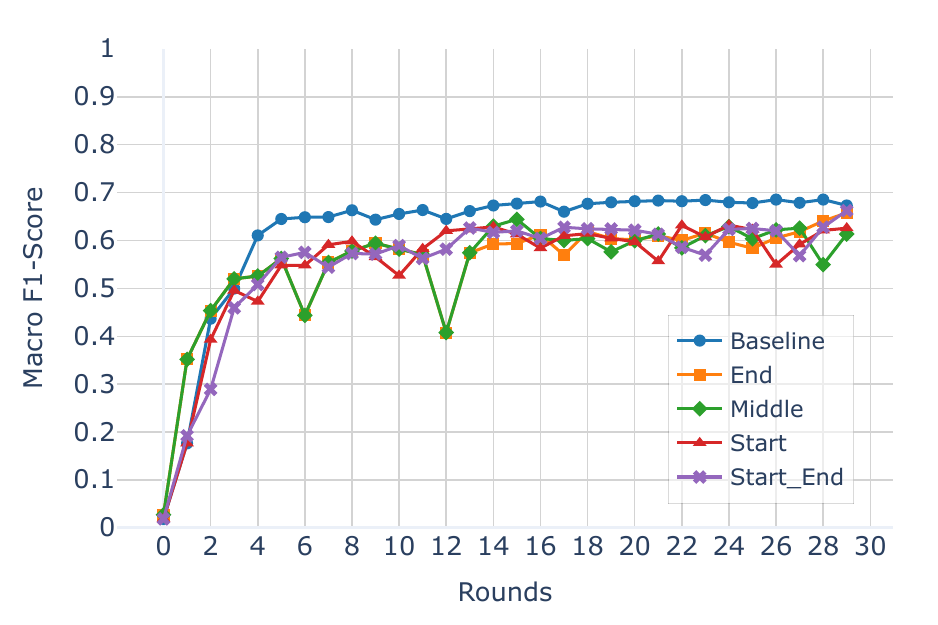}
        \caption{CIFAR10: ResNet}
        \label{fig:availability_impact_cifar10_resnet_bs}
    \end{subfigure}
    \begin{subfigure}{0.48\textwidth}
        \centering
        \includegraphics[width=\linewidth]{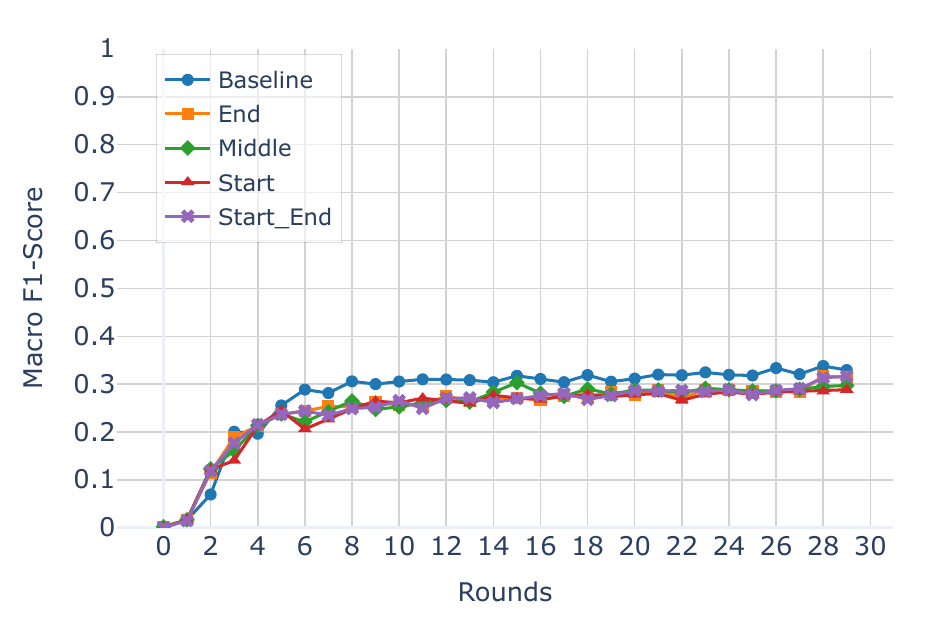}
        \caption{CIFAR100: ResNet}
        \label{fig:availability_impact_cifar100_resnet_bs}
    \end{subfigure}
    \caption{Influence of the ResNet architecture on the availability phases for image datasets with BS. The baseline represents the absence of the failure}
    \label{fig:cifar10_cifar100_timing_resnet}
\end{figure}

\begin{figure}[t]
    \centering
    \begin{subfigure}{0.49\textwidth}
        \centering
        \includegraphics[width=\linewidth]{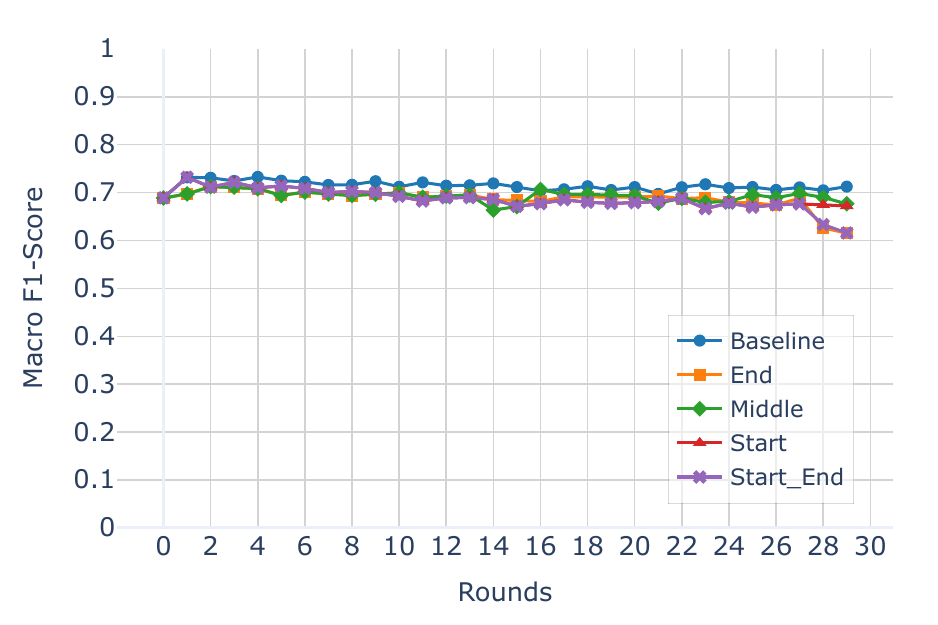}
        \caption{High Skew}
        \label{fig:availability_impact_adult_hs}
    \end{subfigure}
    \begin{subfigure}{0.49\textwidth}
        \centering
        \includegraphics[width=\linewidth]{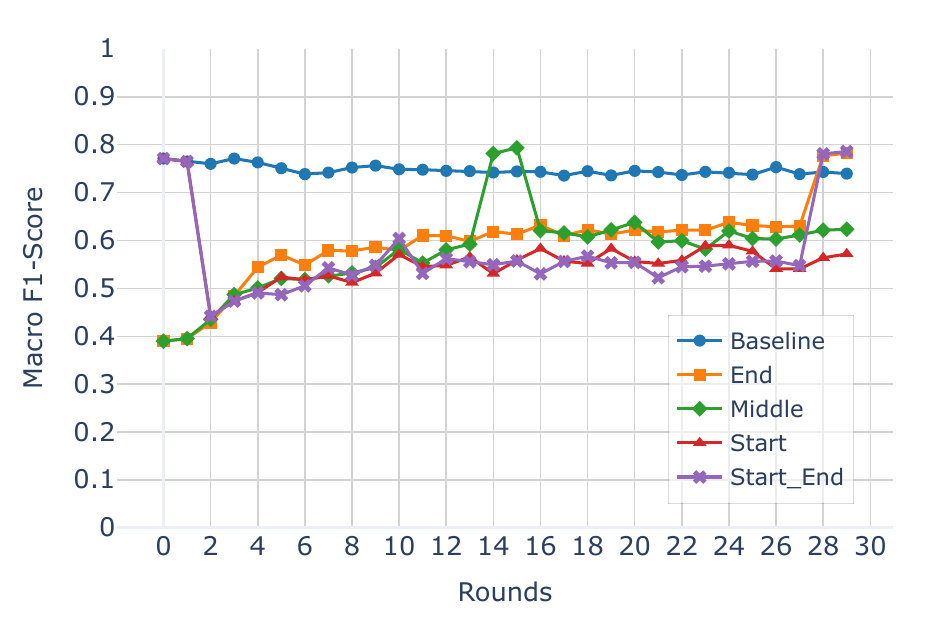}
        \caption{Manual Skew}
        \label{fig:availability_impact_adult_ms}
    \end{subfigure}
    \caption{Adult: Availability phases of the missing participant across HS and MS. The baseline represents the absence of the failure}
    \label{fig:adult_timing}
\end{figure}

\begin{figure}[t]
    \centering
    \begin{subfigure}{0.49\textwidth}
        \centering
        \includegraphics[width=\linewidth]{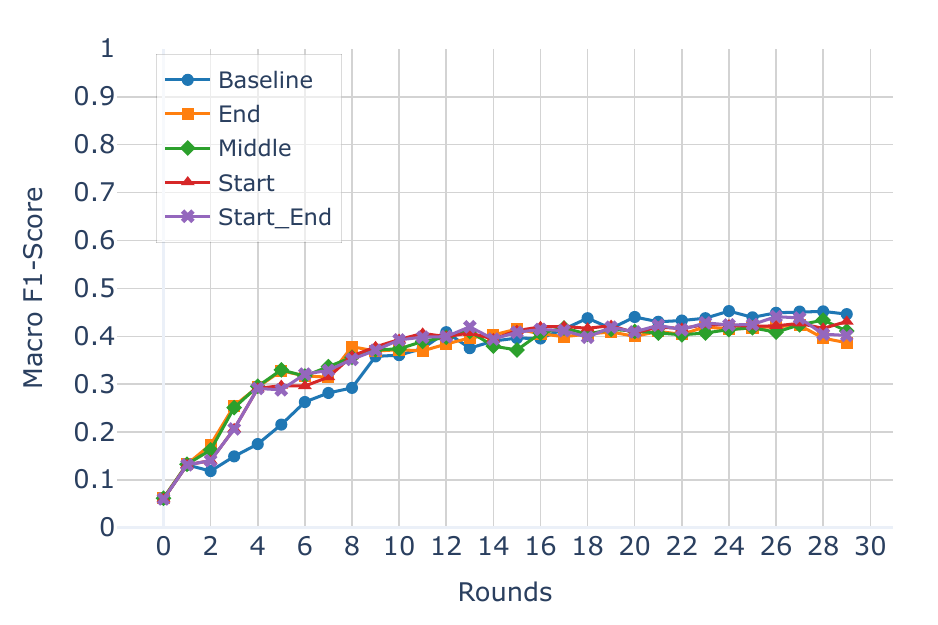}
        \caption{CIFAR10}
        \label{fig:availability_impact_cifar10_manual}
    \end{subfigure}
    \begin{subfigure}{0.49\textwidth}
        \centering
        \includegraphics[width=\linewidth]{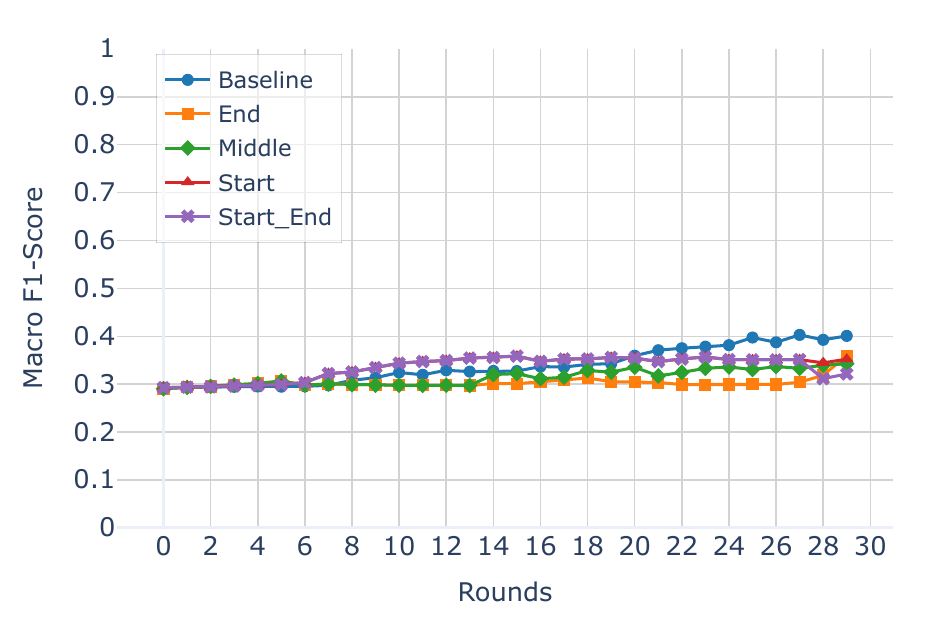}
        \caption{Covertype}
        \label{fig:availability_impact_covertype_manual}
    \end{subfigure}
    \caption{Manuel Skew: Different availability phases for the missing participant across CIFAR10 and Covertype. The baseline represents the absence of the failure}
    \label{fig:datasets_manual}
\end{figure}

\begin{figure}[t]
    \centering
    \begin{subfigure}{0.48\textwidth}
        \centering
        \includegraphics[width=\linewidth]{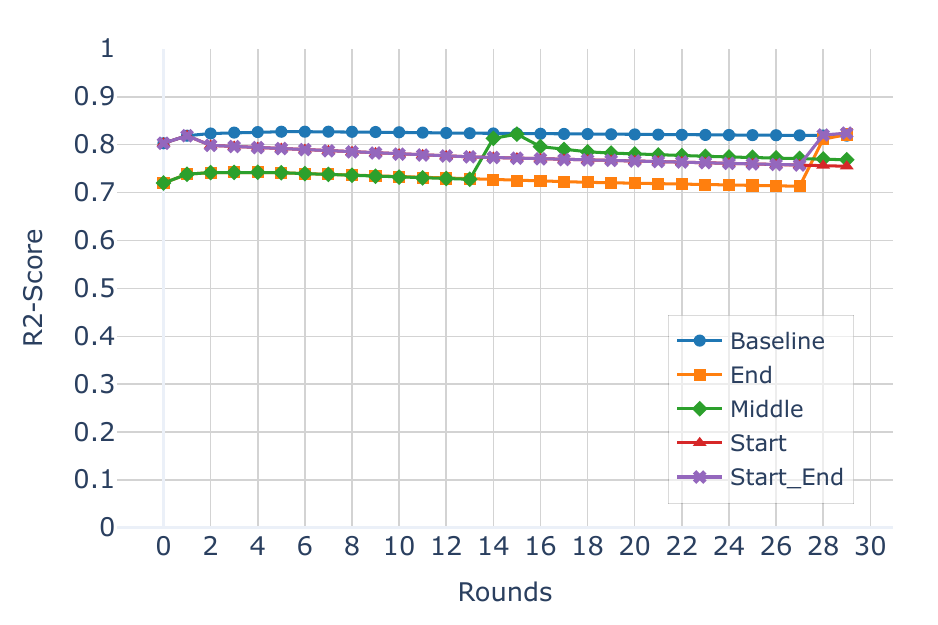}
        \caption{LSTM}
        \label{fig:availability_impact_germansolarfarm}
    \end{subfigure}
    \begin{subfigure}{0.48\textwidth}
        \centering
        \includegraphics[width=\linewidth]{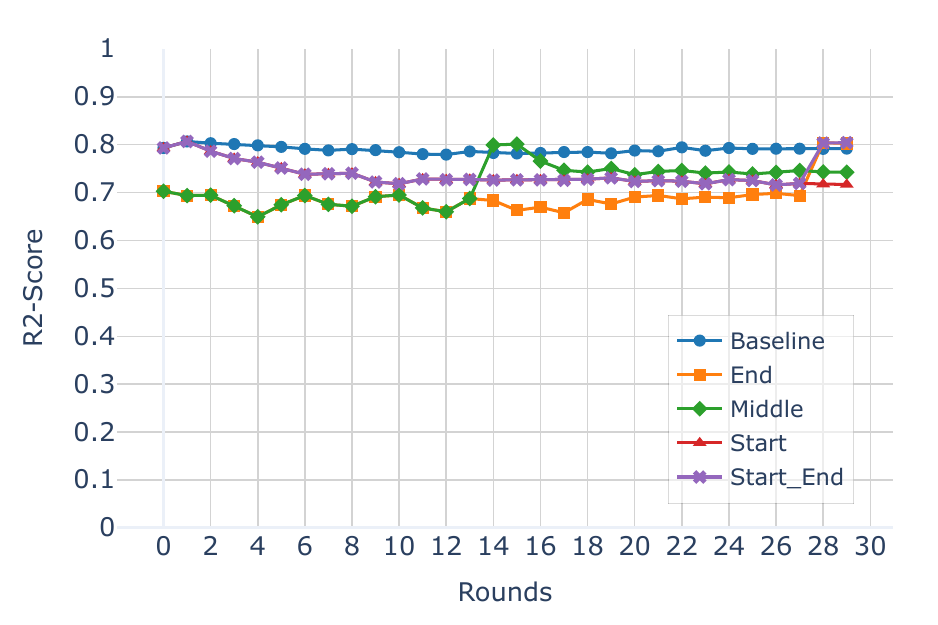}
        \caption{1D-CNN}
        \label{fig:availability_impact_germansolarfarm_cnn}
    \end{subfigure}
    \caption{GermanSolarFarm: Different availability phases for missing participants across different model architectures, where the baseline represents the absence of the failure}
    \label{fig:germansolarfarm_timing}
\end{figure}

\noindent To deduce this finding, we start by comparing the base skew and the high label skew for \textit{CIFAR10} in Fig.~\ref{fig:cifar10_timing}, which shows that in each availability phase, the performance spikes.
Here, the availability at the end of the training shows a performance recovery.
If we combine it with availability at the start, then the recovery improves.
However, it does not match the baseline.
We assume that during the availability phases, the model has learned new patterns, improving the quality of the global model. 
It is also possible that new samples were added that can be predicted by the model very well; however, the F1-score should limit this behavior as it treats all classes equally.
For \textit{CIFAR100}, we can observe similar results.
Furthermore, we analyze the impact of the ResNet architecture in Fig.~\ref{fig:cifar10_cifar100_timing_resnet}.
While the overall performance is fluctuating, we can identify that the model architecture does not change the effect of different availability. Here, the performance spikes for the timing in the middle, which corresponds to rounds 14 and 15.
% Tabular
Next, we analyze the tabular datasets.
% DSkew Skew has high impact, while Base Skew has less impact => DComp + DSkew
Here, for the \textit{Adult} dataset, Fig.~\ref{fig:adult_timing} shows the results for the high and manual skew.
%%%%%%%%%%%%%%%%%%%%%%
\begin{figure}[t]
    \centering
    \begin{subfigure}{0.49\textwidth}
        \centering
        \includegraphics[width=\linewidth]{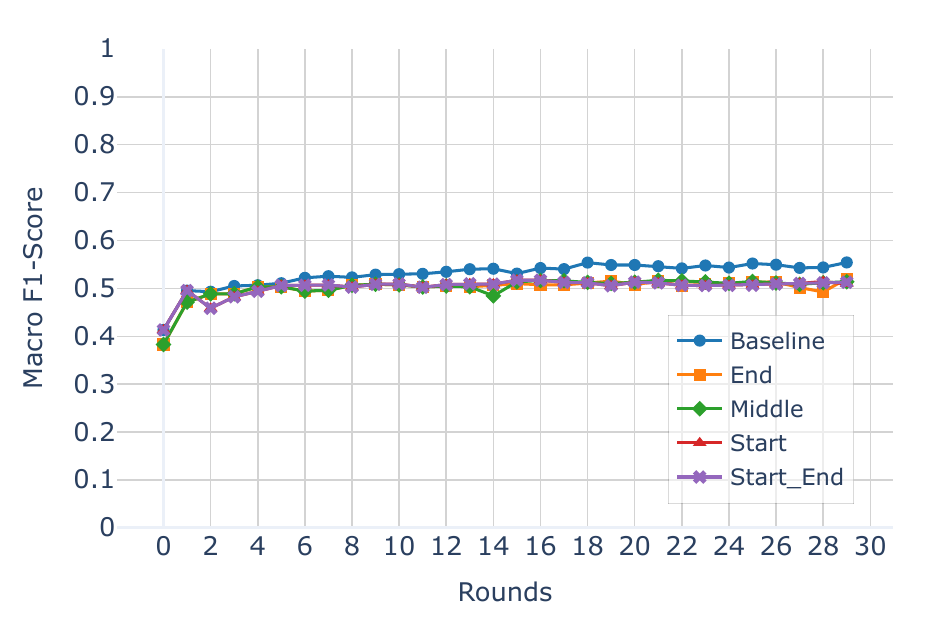}
        \caption{Base Skew}
        \label{fig:availability_impact_covertype_bs}
    \end{subfigure}
    \begin{subfigure}{0.49\textwidth}
        \centering
        \includegraphics[width=\linewidth]{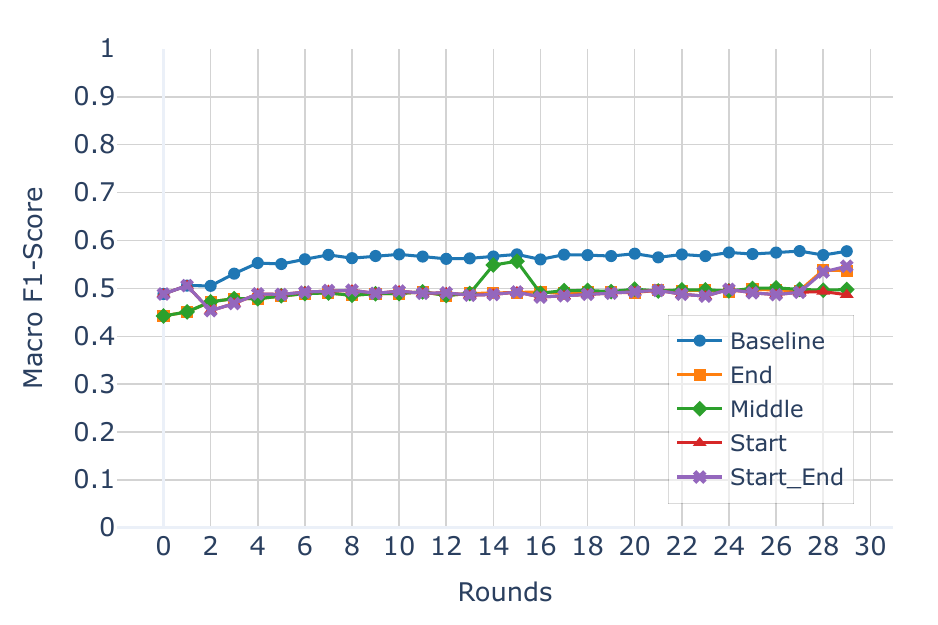}
        \caption{High Skew}
        \label{fig:availability_impact_covertype_hs}
    \end{subfigure}
    \caption{Covertype: Availability phases for the missing participant across the BS and HS. The baseline represents the absence of the failure}
    \label{fig:covertype_timing}
\end{figure}
For the manual skew of \textit{Adult}, which does not consist of unique silos as there are only two classes in total, we can observe the timing shows high importance.
In contrast, the high skew only shows an improvement at the beginning, but in the later rounds the timing leads to a performance reduction.
We explain this behavior by considering that the participant with the most data fails, which will also have the highest impact when rejoining.
Here, the quantity difference is important. 
Furthermore, if we focus on comparing the impact on the quality, we can identify that for the start timing, a higher skew leads to lower performance.
Second, the combination of quantity and labels skew for two classes can lead to participants that have many samples of one class and less for the other class.
In the label distribution, the first participant has mostly labels for the first class, while the other three participants combined have fewer samples for the first class and more for the second one.
Hence, the rejoining can lead to a confusion for the model, which impacts the quality.
% DComp
In the case of \textit{Covertype} in Fig.~\ref{fig:covertype_timing}, the high skew configuration shows a similar impact for timing as in the case of the image datasets.
% \begin{figure}[t]
%     \centering
%     \begin{subfigure}{0.49\textwidth}
%         \centering
%         \includegraphics[width=\linewidth]{Fig13.pdf}
%         \caption{Base Skew}
%         \label{fig:availability_impact_covertype_bs}
%     \end{subfigure}
%     \begin{subfigure}{0.49\textwidth}
%         \centering
%         \includegraphics[width=\linewidth]{Fig14.pdf}
%         \caption{High Skew}
%         \label{fig:availability_impact_covertype_hs}
%     \end{subfigure}
%     \caption{Covertype: Availability phases for the missing participant across the BS and HS. The baseline represents the absence of the failure}
%     \label{fig:covertype_timing}
% \end{figure}
Contrarily, in the base skew, the availability phases barely influence the quality.
We assume that the unbalanced datasets and, hence, the data distribution affect possible timing effects.
To further analyze the impact of DSkew, we compare the manual skew for \textit{CIFAR10}, and \textit{Covertype} in Fig.~\ref{fig:datasets_manual}. 
Here, we can identify that once a class belongs solely to a specific silo and, hence, the overall distribution is extremely skewed, it can lead to a behavior where the timing is not relevant anymore.
Next, we analyze the time-series datasets.
In the case of the \textit{GermanSolarFarm} in Fig.~\ref{fig:germansolarfarm_timing}, the effect is similar to that for the image datasets. 
In addition, we compare the LSTM and the 1-D CNN. 
Similar to previous observations, the LSTM is less fluctuating and seems more robust.
We assume that this is due to LSTMs being designed for remembering long-term dependencies by maintaining states. 
For the \emph{PVOD} datasets, we make the same observation, but the differences are small, as the overall performance loss is small.

Besides the model architecture, we assume that the impact of forgetting is higher in FL as compared to centralized ML, i.e., beyond the number of local training rounds and learning rate, because the aggregation method (e.g., FedAvg) strongly influences the change of model weights.
The findings show that under consideration of the data distribution and potential recovery of a participant, the quality of the model can recover from the impact if the participant rejoins and contributes.
Therefore, this information can help ML engineers to make better decisions on how to handle participant recovery.
Furthermore, it indicates that a missing participant in a skewed setting can lead to a reduced quality.

\subsection{RQ3: How does a participant failure affect the usability of the trained model for the missing participant?}
\begin{figure}[t]
    \centering
    \begin{subfigure}{0.48\textwidth}
        \centering
        \includegraphics[width=\linewidth]{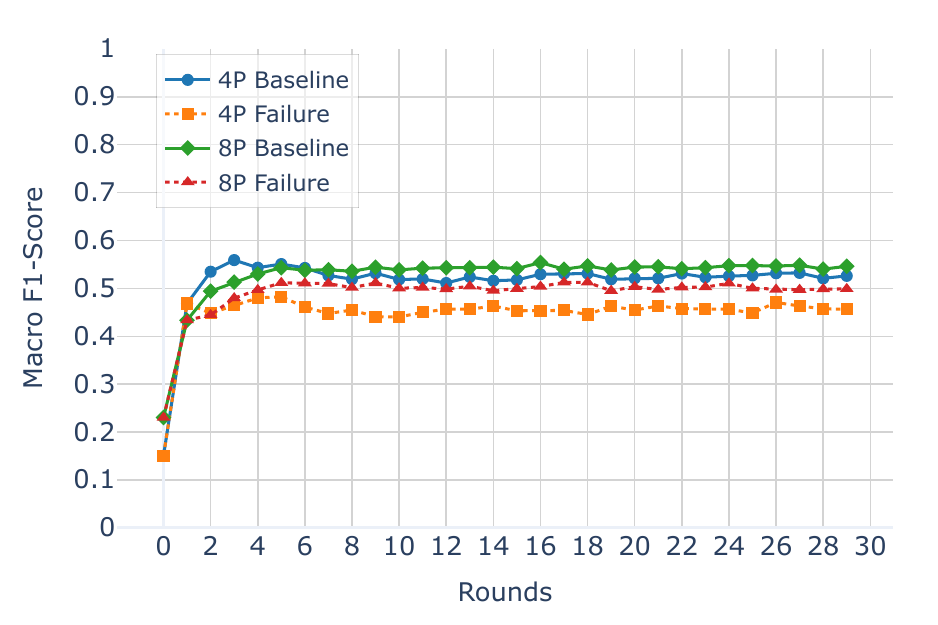}
        \caption{CIFAR10}
        \label{fig:inc_participants_cifar10_bs}
    \end{subfigure}
    \begin{subfigure}{0.48\textwidth}
        \centering
        \includegraphics[width=\linewidth]{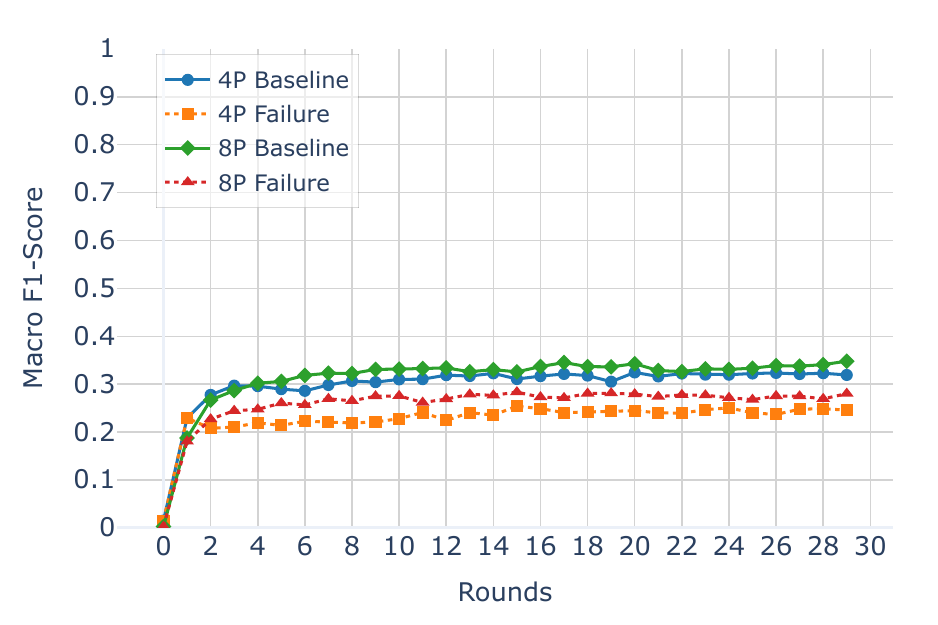}
        \caption{CIFAR100}
        \label{fig:inc_participants_cifar100_bs}
    \end{subfigure}
    \caption{Image BS: Comparison of the model performance for the missing participant during failure and participation (baseline) across different numbers of participants}
    \label{fig:increase_participants_image}
\end{figure}

\begin{figure}[t]
    \centering
    \begin{subfigure}{0.48\textwidth}
        \centering
        \includegraphics[width=\linewidth]{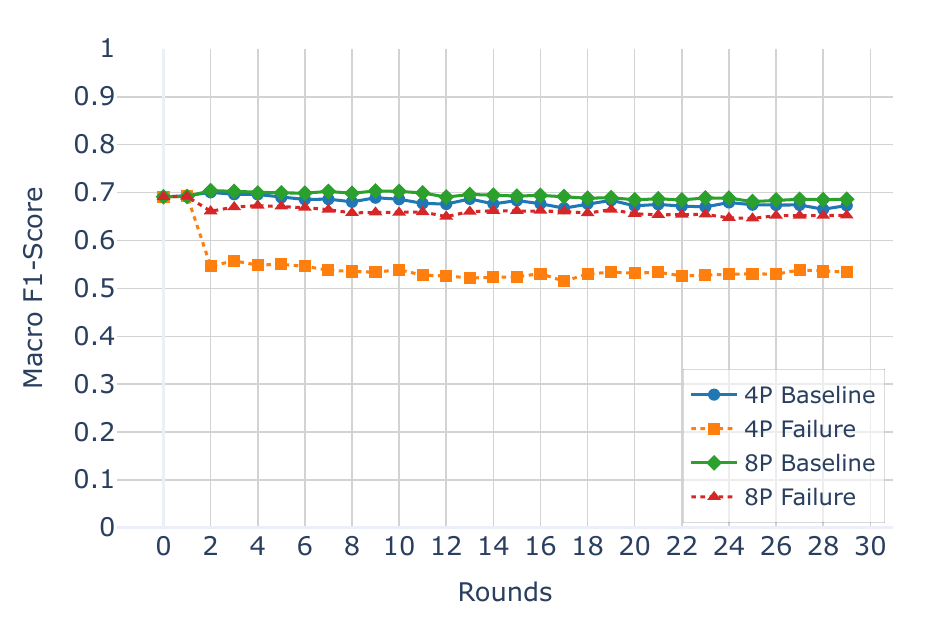}
        \caption{Adult}
        \label{fig:inc_participants_adult_bs}
    \end{subfigure}
    \begin{subfigure}{0.48\textwidth}
        \centering
        \includegraphics[width=\linewidth]{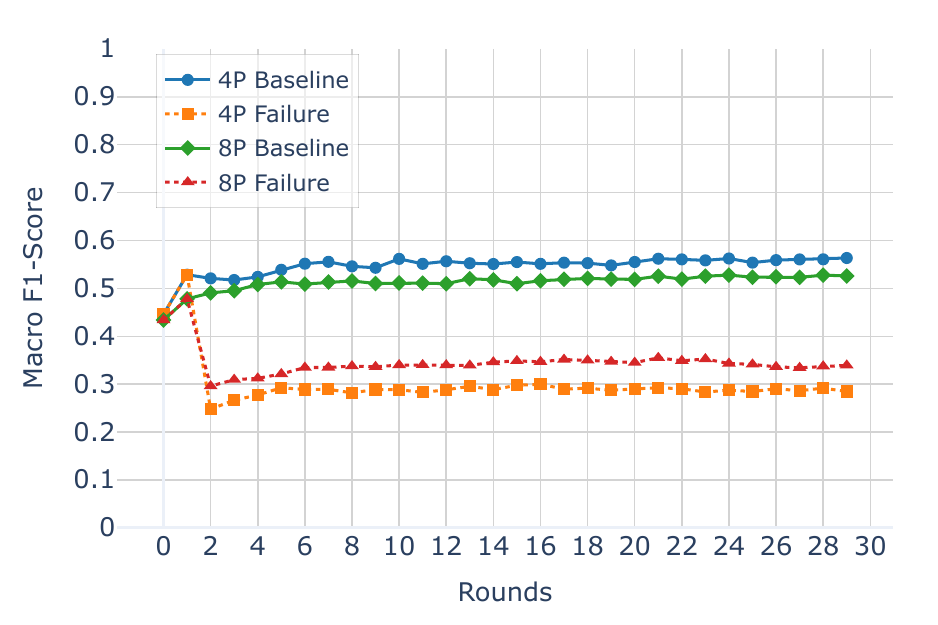}
        \caption{Covertype}
        \label{fig:inc_participants_covertype_bs}
    \end{subfigure}
    \caption{Tabular BS: Comparison of the model performance for the missing participant during failure and participation (baseline) across different numbers of participants}
    \label{fig:increase_participants_tabular}
\end{figure}

\begin{figure}[t]
    \centering
    \begin{subfigure}{0.48\textwidth}
        \centering
        \includegraphics[width=\linewidth]{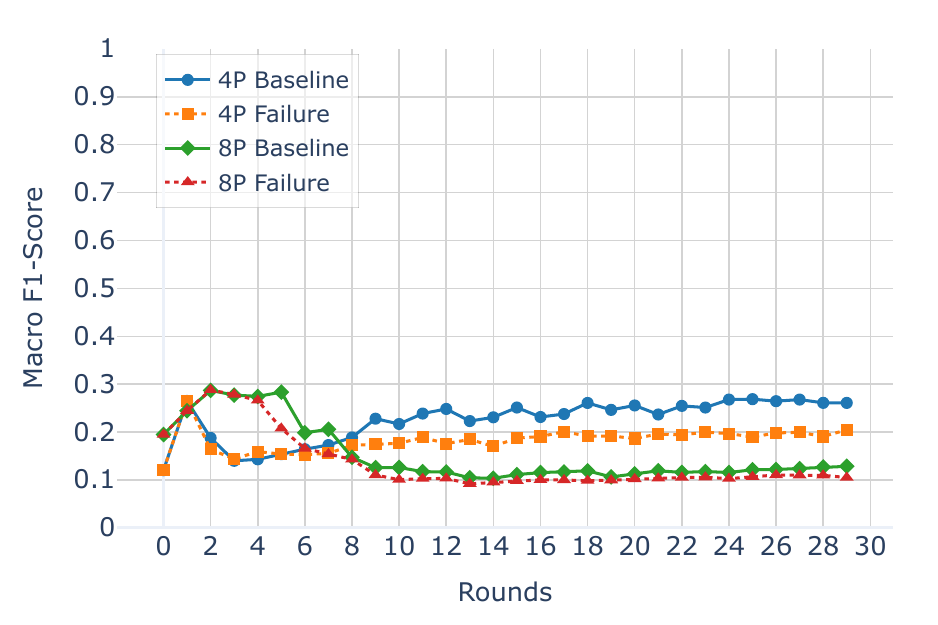}
        \caption{CIFAR10}
        \label{fig:increase_participants_cifar10_ms}
    \end{subfigure}
    \begin{subfigure}{0.48\textwidth}
        \centering
        \includegraphics[width=\linewidth]{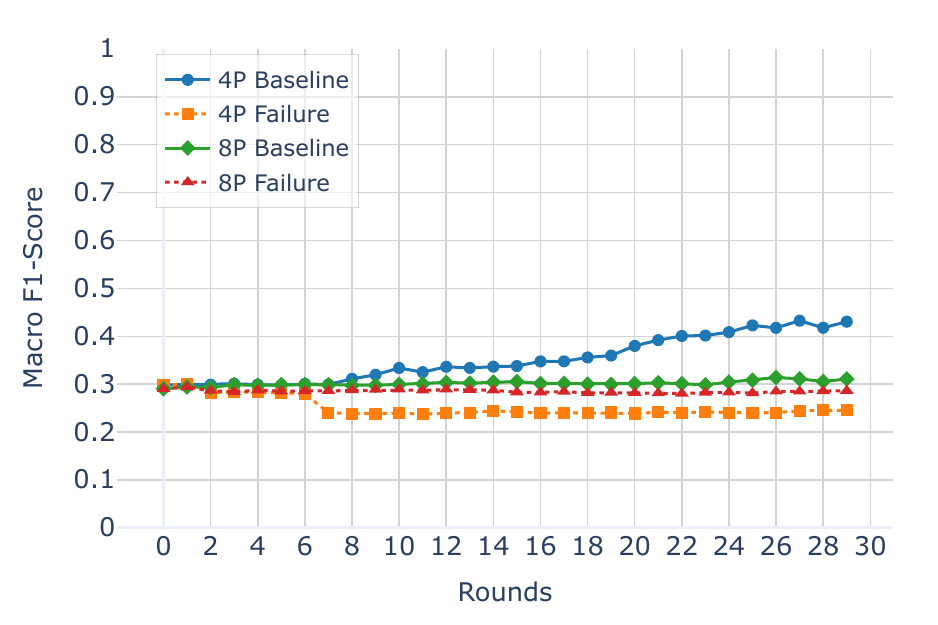}
        \caption{Covertype}
        \label{fig:inc_participants_covertype}
    \end{subfigure}
    \caption{MS: Comparison of the global model performance for the missing participant during failure and participation (baseline) across different numbers of participants}
    \label{fig:increase_participants_manual}
\end{figure}

\begin{mdframed}
\textbf{Experiments:} We examine the performance change for the missing participants by comparing a no-failure against rejoining post-training scenario, while considering the modifiers DSkew, DComp, and NParti.

\noindent\textbf{Finding:} \emph{The failed participant can use the global model with decreased performance. Additional participants with similar data improve usability, while those with new, unique data lead to a decrease.}
Here, the  modifiers DSkew, DComp, and NParti have a strong influence on the model performance for the missing participant as they define the data distributions.
In the skewed settings, additional participants that do not provide similar information can lead to a significant performance drop.
While similar information can compensate for the loss of information across all data modalities.
\end{mdframed}
We start by analyzing Fig.~\ref{fig:increase_participants_image}~and~\ref{fig:increase_participants_tabular}. Here, we can see that the performance of the global model for the missing participant drops significantly.
In addition, we can also observe that the usability of the model increases with more participants.
\begin{figure}[t]
    \centering
    \begin{subfigure}{0.48\textwidth}
        \centering
        \includegraphics[width=\linewidth]{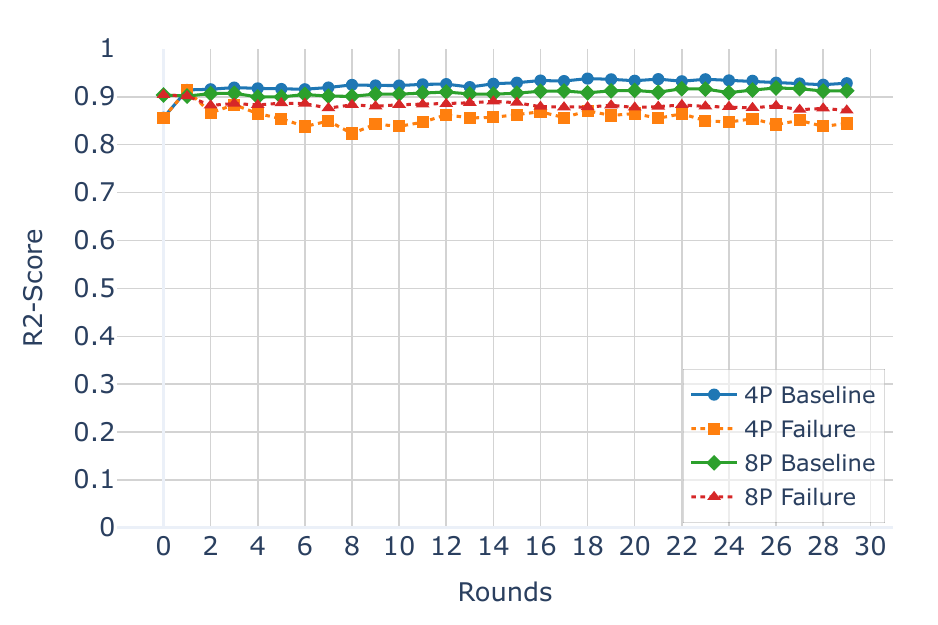}
        \caption{PVOD}
        \label{fig:inc_participants_pvod}
    \end{subfigure}
    \begin{subfigure}{0.48\textwidth}
        \centering
        \includegraphics[width=\linewidth]{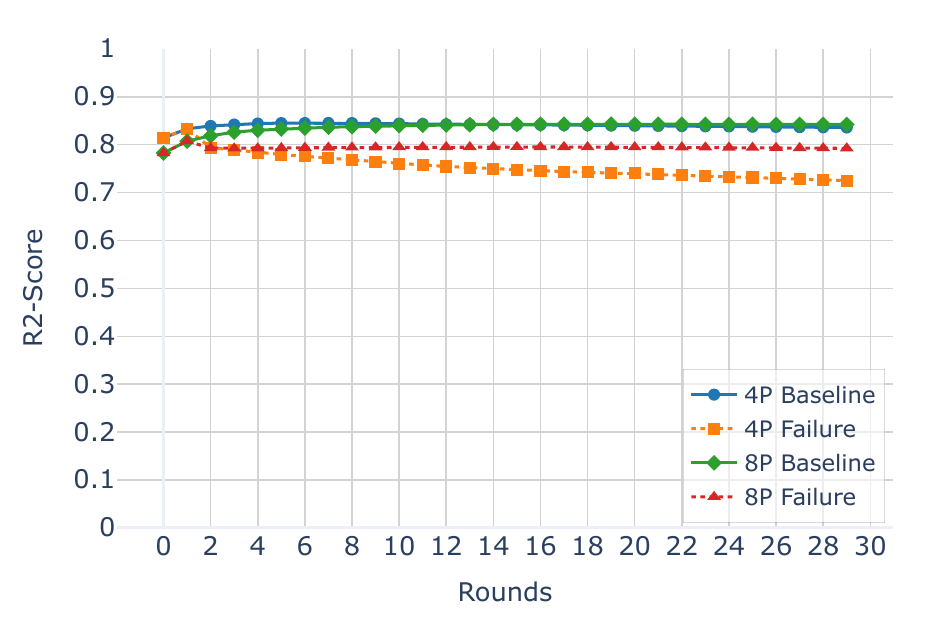}
        \caption{GermanSolarFarm}
        \label{fig:inc_participants_germansolarfarm}
    \end{subfigure}
    \caption{Time-series: Comparison of the model performance for the missing participant during failure and participation (baseline) across different numbers of participants}
    \label{fig:increase_participants_timeseries}
\end{figure}
In Fig.~\ref{fig:increase_participants_timeseries}, we can identify that for \textit{GermanSolarFarm}, the applicability of the global model improves with additional participants. We can make the same observation for \textit{PVOD}.
To further understand how additional participants affect the performance, we analyze DSkew.
For \textit{CIFAR10} in Fig.~\ref{fig:increase_participants_cifar10_ms}, we can identify that unique data cause a reduction in the quality, with additional participants not providing new information but introducing new unique classes to the process.
We can observe the same behavior for \textit{Covertype} in Fig.~\ref{fig:inc_participants_covertype}.
Here, the increase in performance with more participants is small compared to the behavior of the high skew in Fig.~\ref{fig:inc_participants_covertype_bs}. Hence, highly skewed and new unique classes introduced by participants can negatively affect the performance.
When introducing similar information already in the training process, the usability of the model for the missing participant can be improved. The \textit{Adult} dataset with fewer classes shows in Fig.~\ref{fig:inc_participants_adult_bs} that additional participants can improve the usability of the global model for the missing participant. Here, also the modifier DComp influences the skew.
As consequence, it is important to analyze the currently available information as well as new information introduced to the FL process by additional participants to understand how the quality is influenced.

\subsection{RQ4: Do participant contribution measures reflect the impact of a missing participant on the training task?}
\begin{figure}[t]
    \centering
    \begin{subfigure}{0.43\textwidth}
        \centering
        \includegraphics[width=\linewidth]{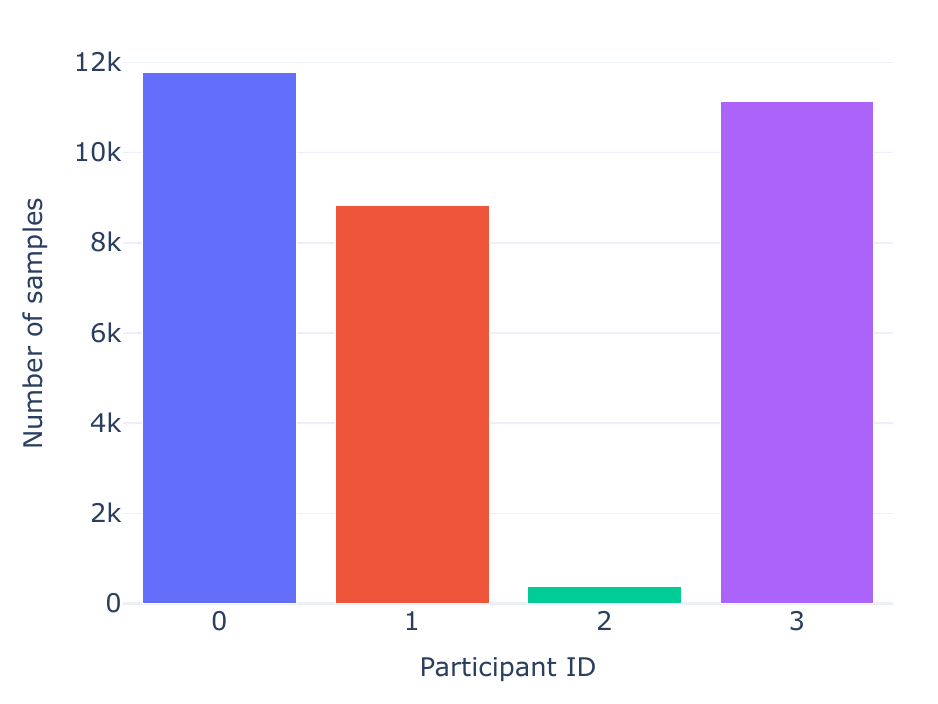}
        \caption{Base Skew}
        \label{fig:cifar10_sv_bs_data_dist}
    \end{subfigure}
    \begin{subfigure}{0.43\textwidth}
        \centering
        \includegraphics[width=\linewidth]{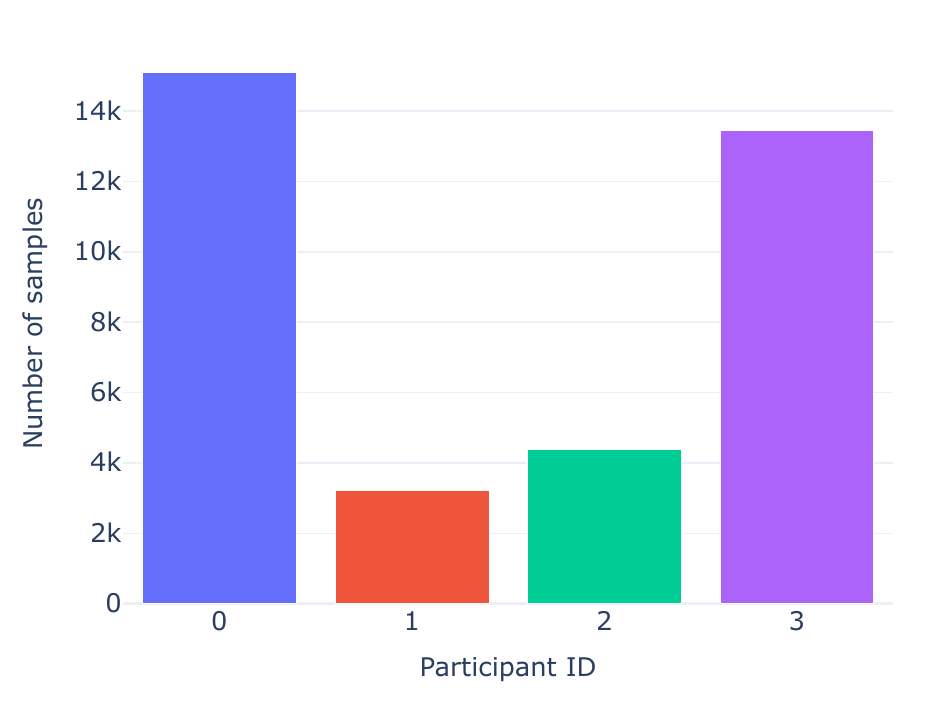}
        \caption{High Skew}
        \label{fig:cifar10_sv_hs_data_dist}
    \end{subfigure}
    \caption{\textit{CIFAR10}: Quantity distribution per participant across different skews}
    \label{fig:cifar10_ld_sv}
\end{figure}

\begin{figure}[t]
    \centering
    \begin{subfigure}{0.48\textwidth}
        \centering
        \includegraphics[width=\linewidth]{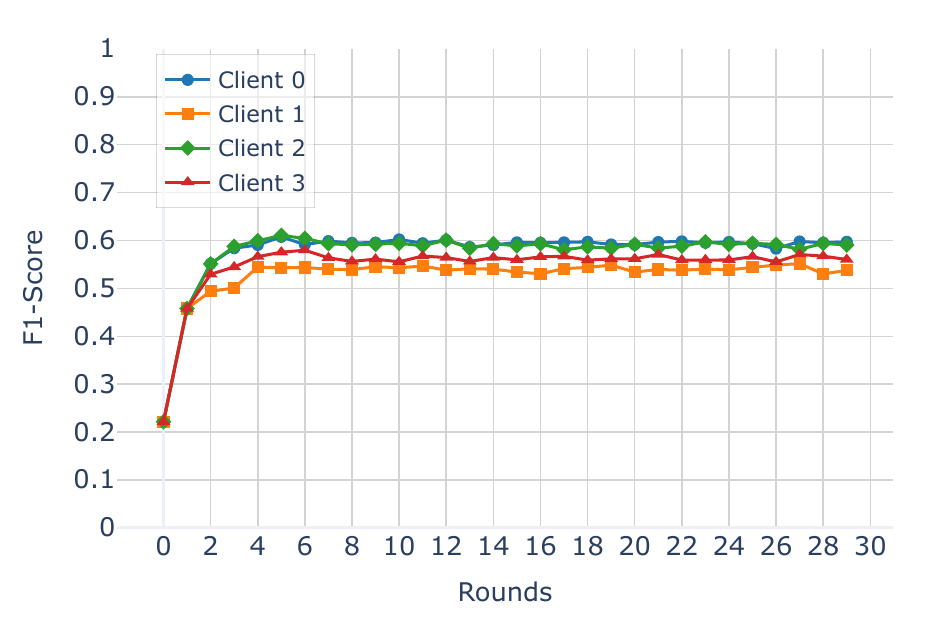}
        \caption{Base Skew}
        \label{fig:cifar10_sv_bs}
    \end{subfigure}
    \begin{subfigure}{0.48\textwidth}
        \centering
        \includegraphics[width=\linewidth]{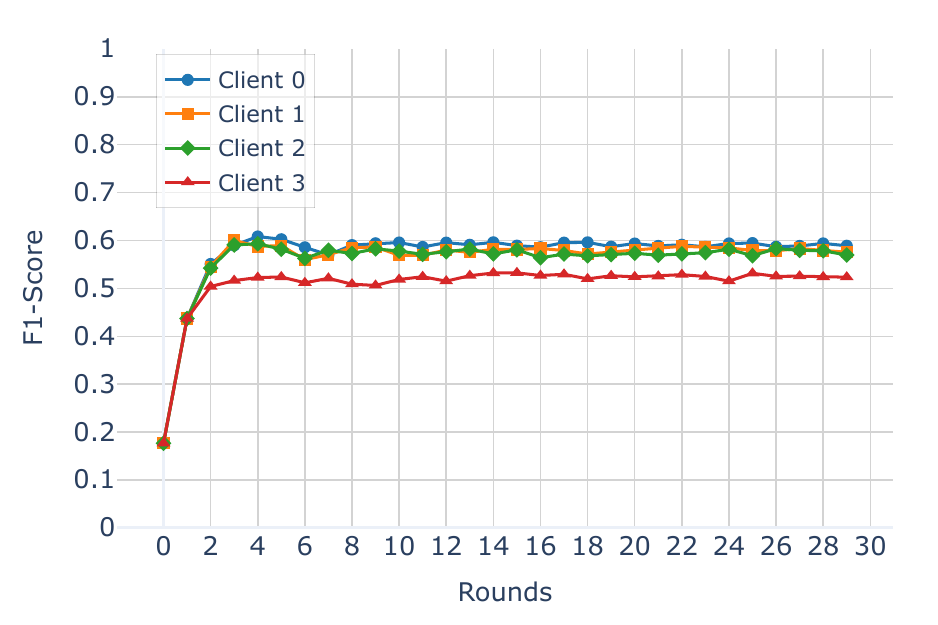}
        \caption{High Skew}
        \label{fig:cifar10_sv_hs}
    \end{subfigure}
    \caption{\textit{CIFAR10}: Impact of different participants dropping out of the process}
    \label{fig:cifar10_sv}
\end{figure}

\begin{mdframed}
\textbf{Experiments:} To analyze the applicability of contribution measures as predictor for the impact, we compare the SVs of a participant with the impact if it is missing.

\noindent\textbf{Finding:} \emph{Participant contribution metrics can detect participants with a very different data distribution compared to the other participants.}
\end{mdframed}
To deduce this answer, we analyze the experiments and the resulting SVs in Table~\ref{tab:cifar_10_shapley_values}.
\begin{table}[t!]
\centering
\caption{\textit{CIFAR10}: SVs for base skew (BS) and high skew (HS) for the first rounds until failure}
\label{tab:cifar_10_shapley_values}
\begin{tabular}{@{}lllll@{}}
\toprule
\textbf{Round} & \textbf{P0-SV} & \textbf{P1-SV} & \textbf{P2-SV}  & \textbf{P3-SV}  \\
\midrule
0-BS &	-0.014305 &	0.039144 & -0.036371 & 0.018734\\
1-BS & -0.004299 &	0.014654 & -0.036120 & 0.031978\\
\midrule
0-HS &	0.045645 &	-0.013814 & 0.000002 & -0.025670\\
1-HS & 0.027890 &	-0.036587 & -0.032981 & 0.049456\\
\bottomrule
\end{tabular}
\end{table}
In combination with the data distribution in Fig.~\ref{fig:cifar10_sv_bs_data_dist}, we can identify that participant 2 is vastly different from the other participants and also has the lowest SVs.
If we compare the SVs with the effect of different participants dropping out in Fig.~\ref{fig:cifar10_sv_bs}, we can identify, that this correlates with the SVs: in the case that participant 2 is missing, the model has the best quality. 
For the other participants, it is difficult to make reliable assumptions, as the order of the SVs does not directly represent the order of the impact.
Next, we compare the \textit{CIFAR10} high skew. The distribution in Fig.~\ref{fig:cifar10_sv_hs_data_dist} shows that both participants 1 and 2 have few samples, while participants 0 and 3 have many samples. The SVs show that participants 1 and 2 are less important than 0 and 3.
In Fig.~\ref{fig:cifar10_sv_hs}, we can see that the behavior aligns with the SVs. 
Despite that, the SVs cannot tell whether participant 3 is more important than participant 0.
We make similar observations for \textit{CIFAR100}.

In summary, this allows ML engineers to determine the relevance of outlier participants with a different data distribution than the other participants. 
Hence, it can help in filtering out participants that are unique and important or have less impact on the quality of the global model.

\section{Discussion} 
\label{sec:validity}

In this section, we discuss aspects that influence the validity of our findings.
First, for predicting the impact of a missing participant, we used SVs to validate whether an assessment is possible.
However, to calculate the SVs, a server-side dataset is required.
In many real-world cases, sending a part of the participants data or using a public dataset is not feasible.
Here, a suitable measurement is necessary that is precise and does not require server-side data.
Second, the extended study considered that a single participant crashes during an FL round, which makes analyzing the impact feasible as the change in the data distribution and the impact are easier to trace.
Despite that, in a real-world setting, it is possible that multiple participants can crash at the same time due to network failures. We assume that for all dataset types the impact will only increase as more missing participants result in more missing data in the process.
Third, analyzing the failure-impact modifiers in complete isolation is not feasible, as each factor (e.g., model architecture, dataset, hyperparameters, aggregation methods) is present while training a model.
Hence, a certain combination of multiple factors is always involved.
Therefore, it is not possible to exclude whether some connections between these factors are causing additional influence. 
Here, it is necessary to define scenarios with base configuration.
Last, the experiments for some modifiers are based on exemplary cases such as the datasets and the model architectures. Transferring the results directly onto other cases is difficult, as they may contain unique properties that were not represented in these cases. Here, future work should consider additional datasets and architectures to validate the generalizability of the results.

\section{Conclusion}\label{sec:conclusion}
In this paper, we conducted an extended study on the impact of participant failures on the model performance in cross-silo FL.
In detail, we systematically derived failure-impact modifiers, including model architecture, dataset types, availability timings, and an increasing number of participants.

We showed that for all dataset types, a missing participant can lead to an optimistic evaluation, which can result in wrong decisions regarding the deployment or retraining of the model. 
Similarly, mechanisms that are executed based on performance thresholds can lead to wrong execution, which influences the reliability of the system.
Furthermore, we provided additional insights on the timing and demonstrated the limitations for highly unique silos.
We showed that the usability of the global model for the missing participant is dependent on the other participants and can be influenced positively or negatively based on dynamic changes in the available data.
In addition, we demonstrated that the model architecture can impact the influence of a failure through different kinds of layers in the model architecture, such as regularization and LSTMs; however, here additional modifiers are relevant to consider.
Overall, the impact of a missing participant is dependent on multiple modifiers that need to be considered jointly in order to obtain a detailed understanding of the failure situation in cross-silo FL.

These insights are essential to understand the impact of missing participants in cross-silo FL.
They demonstrate that, besides the training also the evaluation must be reliable as they are connected in deployment decisions.
The derived modifiers and their influence can help in designing new cost-effective fault-tolerance methods that adapt to different environments of cross-silo FL systems by analyzing the failure-impact modifiers and adapting accordingly.
By leveraging these insights, the robustness and fault-tolerance of FL Systems can be improved, which supports the adoption in real-world challenges.
As future work, we plan to design fault-tolerance methods that use these insights to increase the dependability of FL systems.

\bibliographystyle{plain}
\bibliography{bib}

\end{document}